\documentclass{aa}  
\usepackage{graphicx}
\usepackage{txfonts}
\usepackage{subcaption}
\usepackage{lscape}
\usepackage{placeins} 

\begin{document}

   \title{Revisiting self-interacting dark matter with DESI DR2 baryon acoustic oscillation observations}

   \author{Yixuan Zhu\inst{1}
         \and
         Shuo Cao\inst{1,2}\thanks{\email{caoshuo@bnu.edu.cn}} 
         \and
         Xuanni chen\inst{1}
         \and
         Xuchen Tai\inst{1}
         \and
         Xinyang Jin\inst{1}
         \and
         Yuanming Feng\inst{1}
         }

   \institute{School of Physics and Astronomy, Beijing Normal University, Beijing 100875, China;
         \and
   Institute for Frontiers in Astronomy and Astrophysics, Beijing Normal University, Beijing 102206, China;
     }

   \date{Received XXX/ Accepted XXX}

  \abstract
    {}
   {In this paper, we revisit the self-interacting dark matter (SIDM) scenario, which was originally proposed to resolve the small-scale problems of the standard cold dark matter (CDM) paradigm and additionally provides a mechanism for driving late-time cosmic acceleration.}
   {We utilized the latest DESI baryon acoustic oscillation data, along with type Ia supernova data from Pantheon+/DESY5 and Hubble parameter data from cosmic chronometer measurements, to constrain the SIDM models. Specifically, we considered three representative SIDM scenarios that include the creation and annihilation processes of SIDM particles.}
   {We focus on the annihilation of dark matter particles in model SIDM1; our findings indicate the effective annihilation cross section is $\log_{10}(\kappa C_1)=-4.32$ (with an upper limit of $\log_{10}(\kappa C_1)=-3.90$). For the SIDM2 model, the SIDM fluid behaves similarly to a dark energy component with a mildly dynamical equation of state ($w_{\rm{SIDM}}=-0.82 \pm 0.04$). We also quantitatively discuss the dark matter self-interactions in a more general SIDM model (SIDM3), which reveals a strong correlation between the annihilation cross section and equation of state of dark matter. For the three SIDM models, the measured Hubble constant agrees with that derived from the recent Planck 2018 results; both are inconsistent with the results of the SH0ES collaboration, who report a significantly higher $H_0$ value. The Bayesian evidence indicates that SIDM2 and SIDM3 are more supported by the available observational data than the concordance standard $\Lambda$CDM. Furthermore, the cosmological constraints on the effective annihilation parameter can be translated into bounds on the velocity-averaged annihilation cross section and the self-interaction cross section per unit mass. While low-redshift data alone imply a relatively large cross section, the inclusion of cosmic microwave background distance priors significantly tightens the constraint at cluster-scale velocities, highlighting the necessity of a velocity-dependent cross section to reconcile cosmological and astrophysical observations.}
   {}


   \maketitle

\section{Introduction}\label{sec:1}

Cosmic acceleration, which was first discovered through observations of distant type Ia supernovae (SNe Ia; \citealt{1998AJ....116.1009R, 1999ApJ...517..565P}), has remained one of the most profound mysteries in modern cosmology for over two decades. The standard cosmological model, known as the $\Lambda$ cold dark matter (CDM) model, attributes this acceleration to a cosmological constant, $\Lambda$, which acts as a form of dark energy with a constant equation of state (EoS), $w=-1$. The $\Lambda$CDM model has been remarkably successful in describing the large-scale structure and evolution of the Universe and is supported by a wide array of cosmological observations, including measurements of the cosmic microwave background (CMB) anisotropies \citep{2020A&A...641A...6P}, baryon acoustic oscillations (BAOs; \citealt{2025JCAP...02..021A}), and supernovae light curves \citep{2022ApJ...938..110B}. On galactic and sub-galactic scales, however, the standard CDM model faces several well-known challenges, including the core-cusp problem \citep{2017ARA&A..55..343B}, the missing satellites problem \citep{1999ApJ...522...82K}, and the too-big-to-fail problem \citep{2012MNRAS.422.1203B}. Self-interacting dark matter (SIDM) has emerged as a promising solution to these discrepancies by allowing dark matter (DM) particles to undergo elastic scattering off one another \citep{2000PhRvL..84.3760S, 2001ApJ...547..574D}. By introducing a scattering cross section per unit mass ($\sigma/M_{\chi}$), SIDM can modify structure formation on dwarf and galactic scales while remaining consistent with the successful large-scale predictions of the $\Lambda$CDM model \citep{2024EPJST.233.2113D}. SIDM with velocity-dependent self-interactions has attracted growing attention because it can simultaneously accommodate observational constraints on the scattering cross section per unit mass across a wide range of scales \citep{2020PhRvL.125k1105Y,2021PhRvD.103b3017H}. \citet{2022A&A...659L...5C} proposed a novel method for probing the distribution of DM that exhibits elastic and velocity-independent self-interactions, which can be revealed through multiple measurements of strongly lensed gravitational waves \citep{2021MNRAS.502L..16C}. Recent analysis has revealed the unexpectedly stronger large-scale clustering of diffuse dwarf galaxies, which can be naturally explained by SIDM \citep{2025Natur.642...47Z}. While SIDM was originally conceived to address small-scale structure issues, it also offers the intriguing possibility of explaining the observed cosmic acceleration without invoking a separate dark energy component \citep{2001PhRvD..64f3501Z, 2003NJPh....5...85B}. In the Boltzmann formalism, effective negative pressure can arise from a disequilibrium between particle creation and annihilation processes. Analytic solutions for simple SIDM models indicate that the effective annihilation term is typically small compared to estimates from more general approaches \citep{2009A&A...507...47B}. Notably, the gravitational matter creation model \citep{2008CQGra..25t5006L} is mathematically equivalent to one particular case of SIDM.

Recently, the Dark Energy Spectroscopic Instrument (DESI) survey released its second data release (DR2), which presents BAO measurements from galaxy, quasar, and Lyman-$\alpha$ (Ly$\alpha$) forest tracers \citep{2025PhRvD.112h3514A, 2025PhRvD.112h3515A}. This provides a robust probe of the cosmic expansion rate and angular diameter distance across a wide redshift range. Combined with the latest supernova data from Pantheon+ and the Dark Energy Survey Year 5 (DESY5), the DESI DR2 BAO measurements have provided possible evidence of dynamical dark energy. In this context, several alternative theoretical frameworks have been proposed beyond the $\Lambda$CDM model, including canonical quintessence scalar fields \citep{1988PhRvD..37.3406R, Wetterich1988a, 1998PhRvL..80.1582C}, noncanonical k-essence theories \citep{2000PhRvL..85.4438A, 2001PhRvD..63j3510A}, and various modified gravity models (see, e.g., \citealt{2012PhR...513....1C} for a review). DESI BAO data, in combination with CMB and supernova datasets, have also been used to constrain various aspects of cosmological physics \citep{2025ApJ...987...58W}. In a related study, we used DESI 2024 BAO data combined with luminosity distances from SNe Ia and $H(z)$ measurements from cosmic chronometers (CCs) to constrain the Hubble constant and cosmic curvature \citep{2025ApJ...978L..33G, 2025ApJ...981L..24L}. In this paper, we focus on the specific SIDM models proposed by \citet{2009A&A...507...47B}, which admit simple analytic solutions for the background expansion history. We considered three representative SIDM scenarios: one that reduces to the $\Lambda$CDM limit ($n=0$, $\kappa\neq0$), another that generalizes to a $w$CDM-like expansion history ($n\neq0$, $\kappa=0$), and a more general scenario with both $n\neq0$ and $\kappa\neq0$. These models are characterized by a small number of free parameters that can be constrained using cosmological observations \citep{2011A&A...529A..61C}. We investigated the SIDM framework in the context of cosmological background expansion using the DESI DR2 BAO measurements together with the latest SN Ia data from Pantheon+ and DESY5 and the most recently revised CC $H(z)$ measurements. We also compared the constraints on the SIDM models with those on the standard $\Lambda$CDM model and a more general SIDM scenario applicable to galactic scales in order to explore the potential connection between small-scale structure and large-scale cosmic acceleration within the SIDM framework. The paper is organized as follows. In Sect.~\ref{sec:2} we introduce the fundamental equations of the SIDM framework. In Sect.~\ref{sec:3} we present the observational datasets, the constraint results, and related discussions. We summarize our conclusions in Sect.~\ref{sec:4}.

\section{Methodology and SIDM models}\label{sec:2}

   In this section, we briefly review the fundamental equations of the SIDM models; further details can be found in \citet{2009A&A...507...47B}. We assumed that the total energy density of the cosmic fluid obeys the collisional Boltzmann equation \citep{Kolb:1990vq}
   \begin{equation}
      \dot{\rho} + 3H\rho + \kappa\rho^2 - \Psi = 0,
      \label{eq:1}
   \end{equation}
   where $\rho$ is the total energy density of the cosmic fluid (encompassing DM, baryons, and any exotic energy component), $\Psi$ is the creation rate of DM particle pairs, and the annihilation parameter $\kappa(\geq0)$ is defined as
   \begin{equation}
      \kappa = \frac{\langle\sigma v\rangle}{M_{\chi}}.
      \label{eq:2}
   \end{equation}
   Here, $\sigma$ is the annihilation cross section, $v$ is the mean particle velocity, and $M_{\chi}$ is the DM particle mass. Relative to the standard fluid equation, the effective pressure term takes the form
   \begin{equation}
      P = \frac{\kappa\rho^2 - \Psi}{3H}.
      \label{eq:3}
   \end{equation}
   The microscopic origin of this effective pressure can be understood through the thermodynamics of open systems \citep{1988PNAS...85.7428P, 1989GReGr..21..767P, 1992PhLA..162..223C}. In the presence of particle pair creation, the Gibbs equation acquires an additional term proportional to the chemical potential. The resulting creation pressure $P_{\rm c} = -\Psi/(3H)$ is negative for $\Psi > 0$ and, in the SIDM framework, competes with the positive annihilation pressure $\kappa\rho^{2}/(3H)$. When $\kappa\rho^2-\Psi<0$, meaning the SIDM particle creation rate exceeds the annihilation rate, SIDM can serve as a source of negative pressure in the global dynamics of the Universe. This mechanism is qualitatively different from both quintessence and modified gravity: it does not introduce a new scalar degree of freedom nor a modification of general relativity; instead it follows directly from the collisional Boltzmann equation describing the phase-space evolution of self-interacting particles. Among the functional forms for which this equation admits analytic solutions \citep{2009A&A...507...47B}, only one is of physical interest -- the unique choice given by
   \begin{equation}
      \Psi(a) = a H(a)R(a) = C_1(n + 3)a^n H(a) + \kappa C_1^2 a^{2n}.
      \label{eq:4}
   \end{equation}
   This form simultaneously satisfies two requirements: the background solution admits a closed-form expression for the Hubble parameter, and the matter sector retains the correct $\propto a^{-3}$ scaling in the limit of negligible interactions. The second admissible form identified by \citet{2009A&A...507...47B} yields a matter density decaying as $a^{-2}$, which is inconsistent with large-scale structure formation. The Boltzmann equation reduces to a Riccati equation when expressed in terms of the scale factor, i.e.,
   \begin{equation}
   \mathrm{d}\rho/\mathrm{d}a=f(a)\rho^2+g(a)\rho+R(a), 
   \end{equation}
   with $f=-\kappa/(a H)$, $g=-3/a$, and $R=\Psi/(a H)$, which is fully solvable only when a particular solution is known \citep{2009A&A...507...47B}. The adopted form of Eq.~(\ref{eq:4}) can also be justified as follows: in the matter-dominated epoch, the cosmic fluid is slowly diluted according to $\rho\sim C_1a^n$ ($n\le0$), so $\propto a^n$ constitutes a particular solution of the Boltzmann equation. Moreover, substituting this equation into Eq.~(\ref{eq:1}) reproduces exactly the creation rate of Eq.~(\ref{eq:4}).

   Now the total energy density is expressed as
   \begin{equation}
      \rho(a) = C_1 a^n + \frac{a^{-3}F(a)}{C_2 - \int_1^{a}x^{-3}f(x)F(x)\,\mathrm{d}x},
      \label{eq:5}
   \end{equation}
   where $f(a)=-\kappa/[a H(a)]$, and the kernel function $F(a)$ is given by
   \begin{equation}
      F(a) = \exp\left[-2\kappa C_1\int_1^{a}\frac{x^{n-1}}{H(x)}\,\mathrm{d}x\right].
      \label{eq:6}
   \end{equation}
   The first term in Eq.~(\ref{eq:5}) represents the energy density associated with residual matter creation arising from a possible disequilibrium between particle creation and annihilation, while the second term corresponds to the energy density of SIDM particles dominated by the annihilation process. The two key parameters in this solution are $n$ and $\kappa$, which govern the effective EoS of the SIDM fluid and the strength of the annihilation process, respectively. Introducing a dimensionless $\Omega$-like parameterization, the global density evolution can be recast as
   \begin{equation}
      \left(\frac{H}{H_0}\right)^2 = \Omega_{1,0}a^n + \frac{\Omega_{1,0}\Omega_{2,0}a^{-3}F(a)}{\Omega_{1,0} + \kappa C_1\Omega_{2,0}\int_1^a\frac{F(x)}{x^4 H(x)}\,\mathrm{d}x},
      \label{eq:7}
   \end{equation}
   where $\Omega_{1,0} = 8\pi GC_1/3H_0^2$ and $\Omega_{2,0} = 8\pi G/3H_0^2C_2$ satisfy the normalization condition $\Omega_{1,0} + \Omega_{2,0} = 1$.

   We next considered two limiting cases of the general solution above: $n=0$ and $\kappa=0$. Setting $n=0$, the global density evolution reduces to
   \begin{equation}
      \rho(a) = C_1 + a^{-3}\frac{e^{-2\kappa C_1(t-t_0)}}{C_2 + \kappa Z(t)},
      \label{eq:8}
   \end{equation}
   where $Z(t) = \int_{t_0}^ta^{-3}e^{-2\kappa C_1(t'-t_0)}\,\mathrm{d}t'$. Equation~(\ref{eq:7}) then simplifies to
   \begin{equation}
      \left(\frac{H}{H_0}\right)^2 = \Omega_{1,0} + \Omega_{2,0}a^{-3}\frac{\Omega_{1,0}e^{-2\kappa C_1(t-t_0)}}{\Omega_{1,0} + \kappa C_1\Omega_{2,0}Z(t)},
      \label{eq:9}
   \end{equation}
   where $\Omega_{1,0}$ and $\Omega_{2,0}$ correspond to $\Omega_{\Lambda}$ and $\Omega_{\rm{m}}$ in the $\Lambda$CDM model, respectively. From Eq.~(\ref{eq:2}), the DM particle mass can be related to $\kappa C_1$ (in units of Gyr${}^{-1}$) via
   \begin{equation}
      M_{\chi}=\frac{3.325\times10^{-15}}{\kappa C_1} \frac{\langle\sigma v\rangle}{10^{-26}}h^2(1-\Omega_{2,0})\,\rm{GeV}.
      \label{eq:10}
   \end{equation}
   Here, $h = H_0 / (100~{\rm km\,s^{-1}\,Mpc^{-1}})$ is the dimensionless Hubble constant. In the limit $\kappa \to 0$, the global density evolution takes the form
   \begin{equation}
      \rho(a) = C_1a^{n} + \frac{1}{C_2}a^{-3},
      \label{eq:11}
   \end{equation}
   and Eq.~(\ref{eq:7}) becomes
   \begin{equation}
      \left(\frac{H}{H_0}\right)^2 = \Omega_{1,0}a^{n} + \Omega_{2,0}a^{-3}.
      \label{eq:12}
   \end{equation}
   The condition $n>-2$ implies that the SIDM model possesses an inflection point, and we defined $w_{\rm{SIDM}}=-1-n/3$ as the EoS of SIDM. This condition also guarantees that the present epoch is located after the inflection point, in complete analogy with the standard dark-energy requirement $w<-1/3$. During the epoch of galaxy formation at high redshifts, the interacting DM term still evolves close to the classical $\propto a^{-3}$ behavior, so the effective component responsible for the accelerated expansion remains nearly homogeneous and does not interfere with the growth of structure. Furthermore, the possible radiation products of the DM interactions contribute negligibly to the global dynamics: their density decays as $a^{-4}$ with an additional exponential suppression toward the past, justifying the treatment of the total density as that of the interacting DM component alone \citep{2009A&A...507...47B}. 

\begin{figure}
      \resizebox{\hsize}{!}{\includegraphics{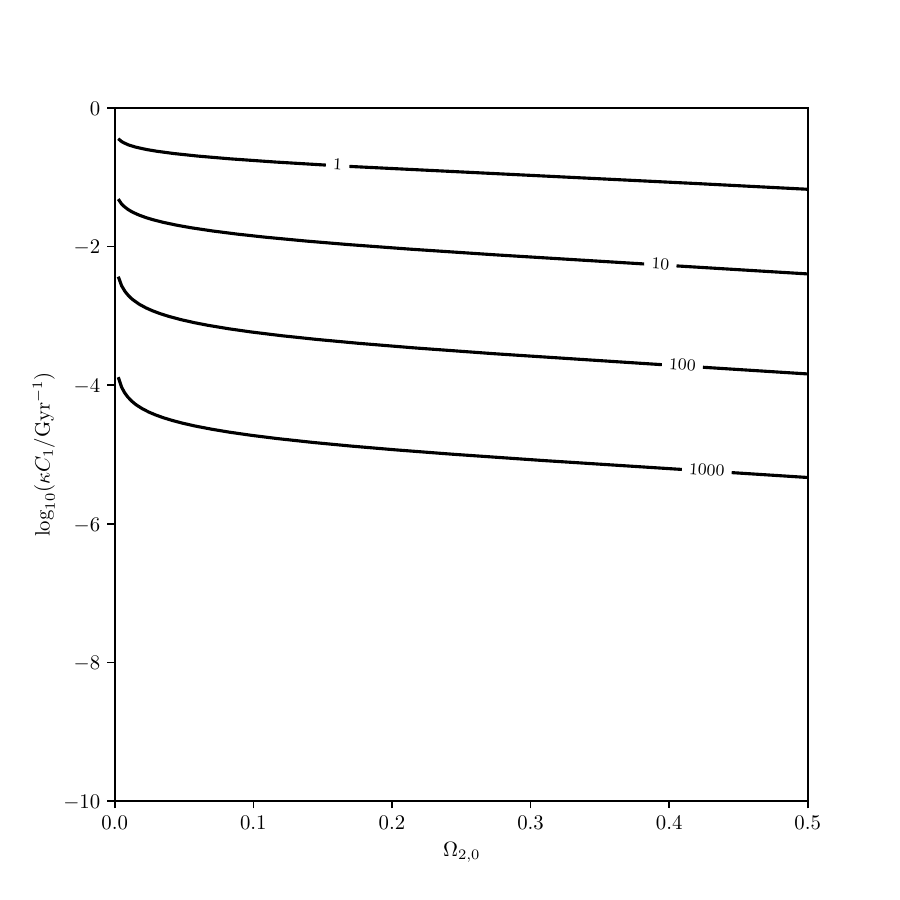}}
      \caption{
         Maximum redshift ($z_{\max}$) as a function of $\Omega_{2,0}$ and $\kappa C_1$, with the Hubble constant fixed at $H_0=68~{\rm km\,s^{-1}\,Mpc^{-1}}$.
      }
      \label{fig:1}
   \end{figure}

   In the following analysis, we therefore considered three representative SIDM scenarios defined as
   \begin{eqnarray}
      {\rm{SIDM1}}: \quad n=0, \kappa\neq0, \nonumber\\
      {\rm{SIDM2}}: \quad n\neq0, \kappa=0, \nonumber\\
      {\rm{SIDM3}}: \quad n\neq0, \kappa\neq0. \nonumber\\
      \nonumber
   \end{eqnarray}
   Note that the $\Lambda$CDM model is mathematically equivalent to the SIDM model with $n=0$ and $\kappa=0$. For all three SIDM models, we related the parameter $\Omega_{2,0}$ to $\Omega_{\rm{m}}$ in $\Lambda$CDM and introduced three free parameters, $\kappa C_1$, $n$, and $H_0$, to quantify the SIDM EoS, the strength of DM particle interactions, and the Hubble constant, respectively. We make some comments on the theoretical solution of SIDM1. The Hubble parameter as a function of redshift ($z$) is defined by
   \begin{equation}
      H(z) = -\frac{1}{1+z}\frac{\mathrm{d}z}{\mathrm{d}t}.
   \end{equation}
   Substituting this equation into Eq.~(\ref{eq:9}) leads to a nonlinear second-order ordinary differential equation (ODE) that governs the redshift evolution:
   \begin{eqnarray}
      2H_0^2\Omega_{1,0}(1+z)^2 z' z'' 
      +\kappa C_1 \left(z'\right)^4-5H_0^2\Omega_{1,0}(1+z)\left(z'\right)^3 \nonumber\\
      +3H_0^4\Omega_{1,0}^2(1+z)^3 z' -H_0^4\Omega_{1,0}^2\kappa C_1\left[(z+1)^4-1\right]\nonumber\\
      =H_0^4\Omega_{1,0}^2\kappa C_1.
   \end{eqnarray}
   Here, primes denote differentiation with respect to cosmic time $t$. We adopted $1+z = \exp(f)$, where $f$ is a function of $t$. The differential equation then reduces to
   \begin{eqnarray}
      2H_0^2\Omega_{1,0}\,f' f''+\kappa C_1 \left(f'\right)^4-3H_0^2\Omega_{1,0}\left(f'\right)^3\nonumber\\
      +3H_0^4\Omega_{1,0}^2\,f'-H_0^4\Omega_{1,0}^2\kappa C_1=0.
   \end{eqnarray}
   Defining $g \equiv f'$, which satisfies $g(t) = -H(t)$, the equation becomes a first-order ordinary differential equation for $g(t)$:
   \begin{equation}
      2H_0^2\Omega_{1,0}\,g g'+\kappa C_1\,g^4-3H_0^2\Omega_{1,0}\,g^3+3H_0^4\Omega_{1,0}^2\,g-H_0^4\Omega_{1,0}^2\kappa C_1=0.
   \end{equation}
   This equation admits an exact analytical solution of the form
   \begin{equation}
      g(t)=\mathcal{G}\left(-\frac{t}{2H_0^2\Omega_{1,0}}+C\right),
   \end{equation}
   where $\mathcal{G}$ denotes the inverse function of $\mathcal{F}(x)$, defined as
   \begin{equation}
      \mathcal{F}(x)=\frac{
         \begin{split}
            3H_0\sqrt{\Omega_{1,0}}\,\mathrm{arctanh}\!\left(\frac{x}{H_0\sqrt{\Omega_{1,0}}}\right)-\kappa C_1\ln\!\left(x^2-H_0^2\Omega_{1,0}\right)\\
            +\kappa C_1\ln\!\left(\kappa C_1 x^2-3H_0^2\Omega_{1,0}x+H_0^2\Omega_{1,0}\kappa C_1\right)
         \end{split}}{9H_0^4\Omega_{1,0}^2-4H_0^2\Omega_{1,0}(\kappa C_1)^2}.
   \end{equation}
   The inverse hyperbolic function is understood in the sense $\mathrm{arctanh}(x)\equiv\frac{1}{2}\ln[(x+1)/(x-1)]$ for $x<-1$, and $C$ is an integration constant fixed by the boundary condition $g(0)\to -\infty$:
   \begin{equation}
      C=\mathcal{F}(-\infty)\equiv\lim_{x\to-\infty}\mathcal{F}(x)=\frac{\kappa C_1\ln\kappa C_1}{9H_0^4\Omega_{1,0}^2-4H_0^2\Omega_{1,0}(\kappa C_1)^2}.
   \end{equation}
   The present cosmic time $t_0$ is then obtained analytically as
   \begin{equation}
      t_0=2H_0^2\Omega_{1,0}\left[\mathcal{F}(-\infty)-\mathcal{F}(-H_0)\right].
   \end{equation}
   The maximum attainable redshift $z_{\max}$ follows from
   \begin{equation}
      z_{\max}=\exp\left[2H_0^2\Omega_{1,0}\int_{-\infty}^{-H_0}\!\!\left(\mathcal{F}(-\infty)-\mathcal{F}(x)\right)\mathrm{d}x+H_0 t_0\right]-1,
   \end{equation}
   which is a function of $\Omega_{2,0}$, $\kappa C_1$, and $H_0$. The dependence of $z_{\max}$ on $\Omega_{2,0}$ and $\kappa C_1$ is illustrated in Fig.~\ref{fig:1}.

\section{Results and discussion}\label{sec:3}

Having defined the SIDM models, we next determined the values of the interaction parameters in each case, along with the remaining cosmological parameters. We employed the publicly available Markov chain Monte Carlo (MCMC) code \texttt{emcee} \citep{2013PASP..125..306F} and analyzed the resulting chains using the \texttt{GetDist} package \citep{2025JCAP...08..025L}. We considered three complementary types of standard cosmological probes to measure angular diameter distances, luminosity distances, and Hubble parameters at different redshifts: DESI BAO data from galaxy, quasar, and Ly$\alpha$ forest tracers; SN Ia data from Pantheon+ and DESY5; and Hubble parameter measurements from CCs.

\subsection{Observational data}\label{sec:3.1}

   The BAO method is a key cosmological probe that is sensitive to the cosmic expansion history and benefits from well-controlled systematics. In this study, we utilized the latest DESI DR2 BAO measurements, which include galaxy, quasar, and Lyman-$\alpha$ tracers spanning the redshift range $[0.295, 2.330]$ across seven distinct redshift bins \citep{2025PhRvD.112h3514A}. Within the flat SIDM cosmological framework, the transverse comoving distance is given by
   \begin{equation}
      D_{\rm{M}}(z) = \frac{c}{H_0}\int_0^z\frac{\mathrm{d}z'}{H(z')/H_0},
   \end{equation}
   and the Hubble distance is defined as
   \begin{equation}
      D_{\rm{H}}(z) = \frac{c}{H(z)}.
   \end{equation}
   The volume-averaged angular diameter distance is expressed as
   \begin{equation}
      D_{\rm{V}}(z) = \left[z D_{\rm{M}}(z)^2 D_{\rm{H}}(z)\right]^{1/3}.
   \end{equation}
   BAO measurements depend on the sound horizon at the drag epoch, $r_d$, which is given by
   \begin{equation}
      r_{\rm{d}} = \int_{z_{\rm{d}}}^{\infty}\frac{c_s(z)}{H(z)}\mathrm{d}z,
   \end{equation}
   where $c_s$ is the sound speed prior to recombination and the drag-epoch redshift $z_{\rm{d}} \approx 1060$. As discussed in Sect.~\ref{sec:2}, the SIDM1 expansion history exhibits a singularity at a maximum redshift $z_{\max}$ below the recombination epoch. Therefore, the sound horizon $r_{\rm d}$, which cannot be computed from first principles within the SIDM1 framework, was treated as a free parameter in the MCMC analysis. Two standardized distance ratios, $D_{\rm{V}}/r_{\rm{d}}$ and $D_{\rm{M}}/D_{\rm{H}}$, are tabulated in \citet{2025PhRvD.112h3515A}. We refer to this full dataset as ``DESI.''

   Type Ia supernovae provided the first evidence of cosmic acceleration and serve as standard candles that directly measure the luminosity distance,
   \begin{equation}
      D_L(z)=(1+z)\int_0^z\frac{c}{H(z')}\,\mathrm{d}z',
   \end{equation}
   with the corresponding distance modulus
   \begin{equation}
      \mu(z) = 5\log_{10}D_L(z) + 25.
   \end{equation}
   We used the most recent compilation of 1550 spectroscopically confirmed SNe Ia from the Pantheon+ survey, which covers the redshift range $[0, 2.26]$ \citep{2022ApJ...938..110B, 2022ApJ...938..113S}, hereafter referred to as ``Pantheon+.'' The Pantheon SN Ia sample has been widely applied to place constraints on different cosmological parameters \citep{2019ApJ...887..163M,2020ApJ...901..129L,2021MNRAS.506.2181L}. We also included the Dark Energy Survey 5-year dataset, which comprises a homogeneously selected sample of 1635 photometrically classified supernovae with redshifts $0.1<z<1.3$, supplemented by 194 low-redshift supernovae with $0.025<z<0.1$, for a total of 1829 SNe \citep{2024ApJ...973L..14D}, hereafter referred to as ``DESY5.''

   The ages of passively evolving early-type galaxies, used as CCs, yield direct measurements of the cosmic expansion rate via
   \begin{equation}
      H(z)=-\frac{1}{1+z}\frac{\Delta z}{\Delta t},
   \end{equation}
   without any prior assumption about the cosmological model. This technique thus provides a cosmology-independent estimate of the Hubble parameter $H(z)$ at different redshifts. A key advantage of the CC approach is that it directly measures $H(z)$ rather than an integral, offering higher sensitivity to cosmological parameters such as the cosmic curvature \citep{CAO2019100274}. This method has also been extensively applied to achieve distance reconstruction independent of any assumptions about cosmology \citep{2017A&A...606A..15C,2021EPJC...81..903L}. In this work, we adopted the recent CC compilation from \citet{2024arXiv241201994M}, which consists of 33 data points spanning the redshift range $[0.07, 1.965]$. We refer to this dataset as ``CC.''

   \begin{table*}
      \caption{Best-fit parameters of the three SIDM models for different dataset combinations.}
      \label{tab:1}
      \centering
      \begin{tabular}{l|ccccc}
         \hline\hline
         Model/Dataset & $\Omega_{2,0}$ or $\Omega_{\rm{m}}$ & $\log_{10}(\kappa C_1/{\rm Gyr}^{-1})$ & $n$ & $w_{\rm{SIDM}}$ & $H_0~[\rm km\,s^{-1}\,Mpc^{-1}]$ \\
         \hline
         $\Lambda$CDM\\
         DESI & $0.298 \pm 0.009$ & --- & --- & --- & --- \\
         DESI+Pantheon+ & $0.314 \pm 0.008$ & --- & --- & --- & --- \\ 
         DESI+DESY5 & $0.314 \pm 0.008$ & --- & --- & --- & --- \\
         DESI+CC+Pantheon+ & $0.314 \pm 0.008$ & --- & --- & --- & $68.4 \pm 1.6$ \\
         \hline
         SIDM1 \\
         DESI & $0.296 \pm 0.010$ & $<-3.59+0.39$ & --- & --- & --- \\
         DESI+Pantheon+ & $0.312 \pm 0.008$ & $<-4.36+0.43$ & --- & --- & --- \\ 
         DESI+DESY5 & $0.314 \pm 0.008$ & $<-4.29+0.40$ & --- & --- & --- \\
         DESI+CC+Pantheon+ & $0.312 \pm 0.008$ & $<-4.32+0.42$ & --- & --- & $68.4 \pm 1.6$ \\
         \hline
         SIDM2 \\
         DESI & $0.297 \pm 0.009$ & --- & $-0.24_{-0.24}^{+0.22}$ & $-0.92_{-0.07}^{+0.08}$ & --- \\
         DESI+Pantheon+ & $0.295 \pm 0.010$ & --- & $-0.55 \pm 0.12$ & $-0.82 \pm 0.04$ & --- \\ 
         DESI+DESY5 & $0.297 \pm 0.009$ & --- & $-0.48 \pm 0.12$ & $-0.84 \pm 0.04$ & --- \\
         DESI+CC+Pantheon+ & $0.296 \pm 0.010$ & --- & $-0.54 \pm 0.12$ & $-0.82 \pm 0.04$ & $66.6 \pm 1.6$ \\
         \hline
         SIDM3 \\
         DESI & $0.289_{-0.005}^{+0.018}$ & $<-2.67 \pm 0.37$ & $-0.27 \pm 0.27$ & $-0.91 \pm 0.09$ & --- \\
         DESI+Pantheon+ & $0.243 \pm 0.070$ & $<-2.12 \pm 0.37$ & $-0.74_{-0.14}^{+0.35}$ & $-0.75_{-0.12}^{+0.05}$ & --- \\ 
         DESI+DESY5 & $0.265_{-0.003}^{+0.051}$ & $<-2.29 \pm 0.35$ & $-0.61_{-0.11}^{+0.27}$ & $-0.80_{-0.09}^{+0.04}$ & --- \\
         DESI+CC+Pantheon+ & $0.263_{-0.006}^{+0.05}$ & $<-2.24 \pm 0.37$ & $-0.67_{-0.09}^{+0.26}$ & $-0.78_{-0.09}^{+0.03}$ & $66.6 \pm 1.5$ \\
         \hline
      \end{tabular}
      \tablefoot{Constraints on the $\Lambda$CDM model are also included for comparison.}
      
   \end{table*}

   \begin{figure}
      \resizebox{\hsize}{!}{\includegraphics{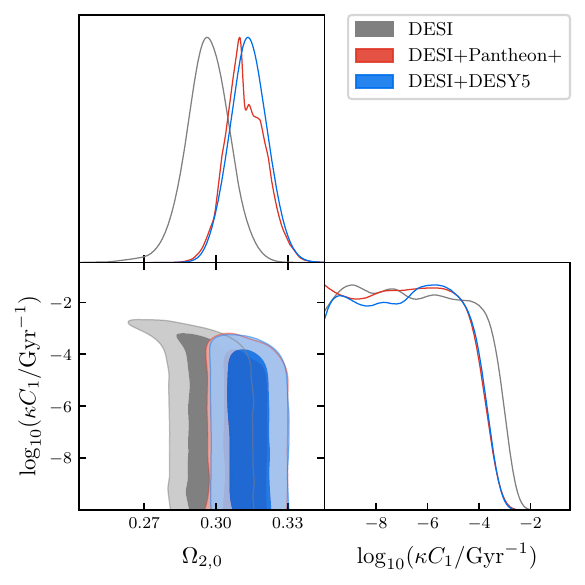}}
      \caption{
         2D marginalized contours and 1D posterior distributions at the $1\sigma$ and $2\sigma$ confidence levels for the cosmological parameters of the SIDM1 model, derived from DESI DR2 combined with Pantheon+/DESY5.
      }
      \label{fig:2}
   \end{figure}

   \begin{figure}
      \resizebox{\hsize}{!}{\includegraphics{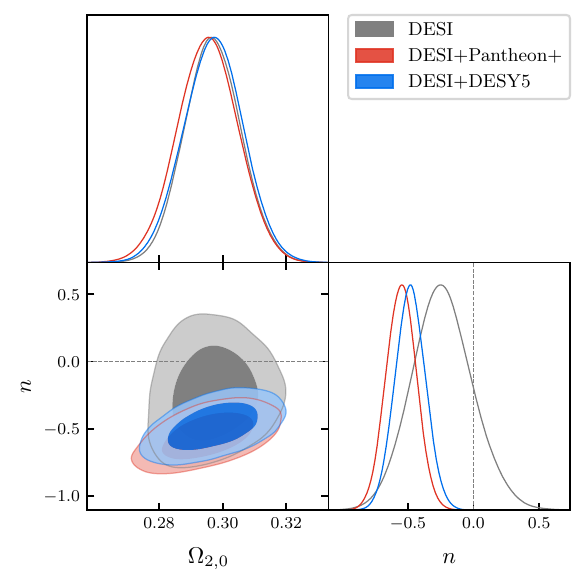}}
      \caption{
         2D marginalized contours and 1D posterior distributions at the $1\sigma$ and $2\sigma$ confidence levels for the cosmological parameters of the SIDM2 model, derived from DESI DR2 combined with Pantheon+/DESY5. The dashed gray lines indicate $n=0$ and $w_{\rm{SIDM}} = -1$.
      }
      \label{fig:3}
   \end{figure}  

   \begin{figure}
      \resizebox{\hsize}{!}{\includegraphics{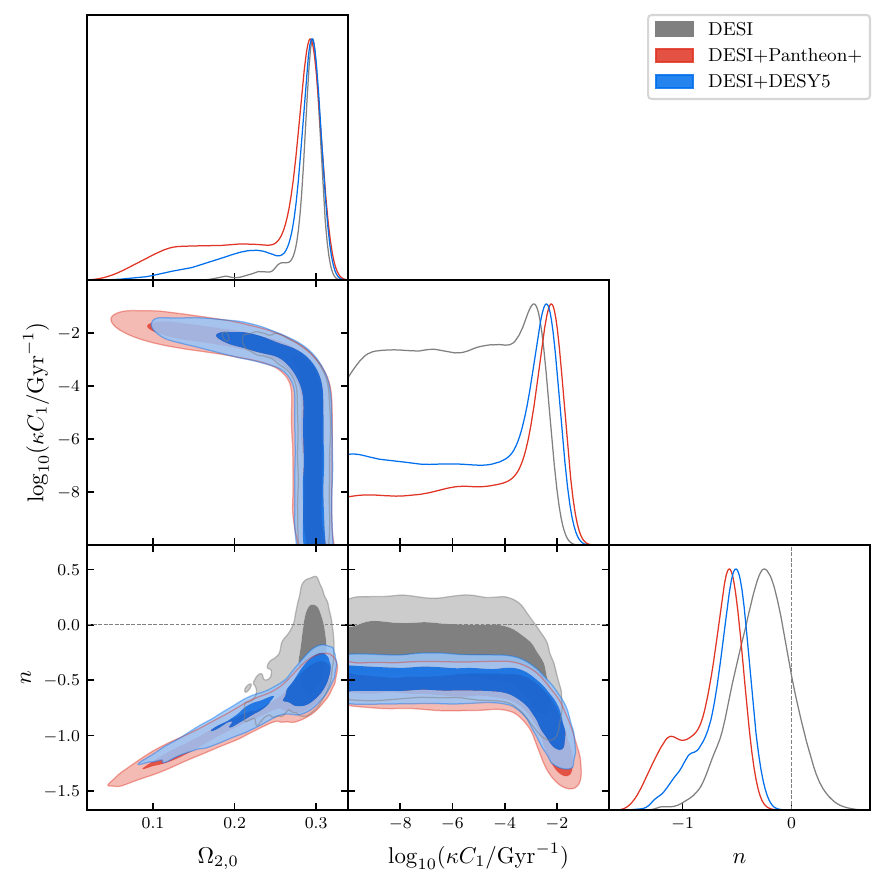}}
      \caption{
         Same as Fig.~\ref{fig:3} but for the SIDM3 model.
      }
      \label{fig:4}
   \end{figure}

   \begin{figure}
      \resizebox{\hsize}{!}{\includegraphics{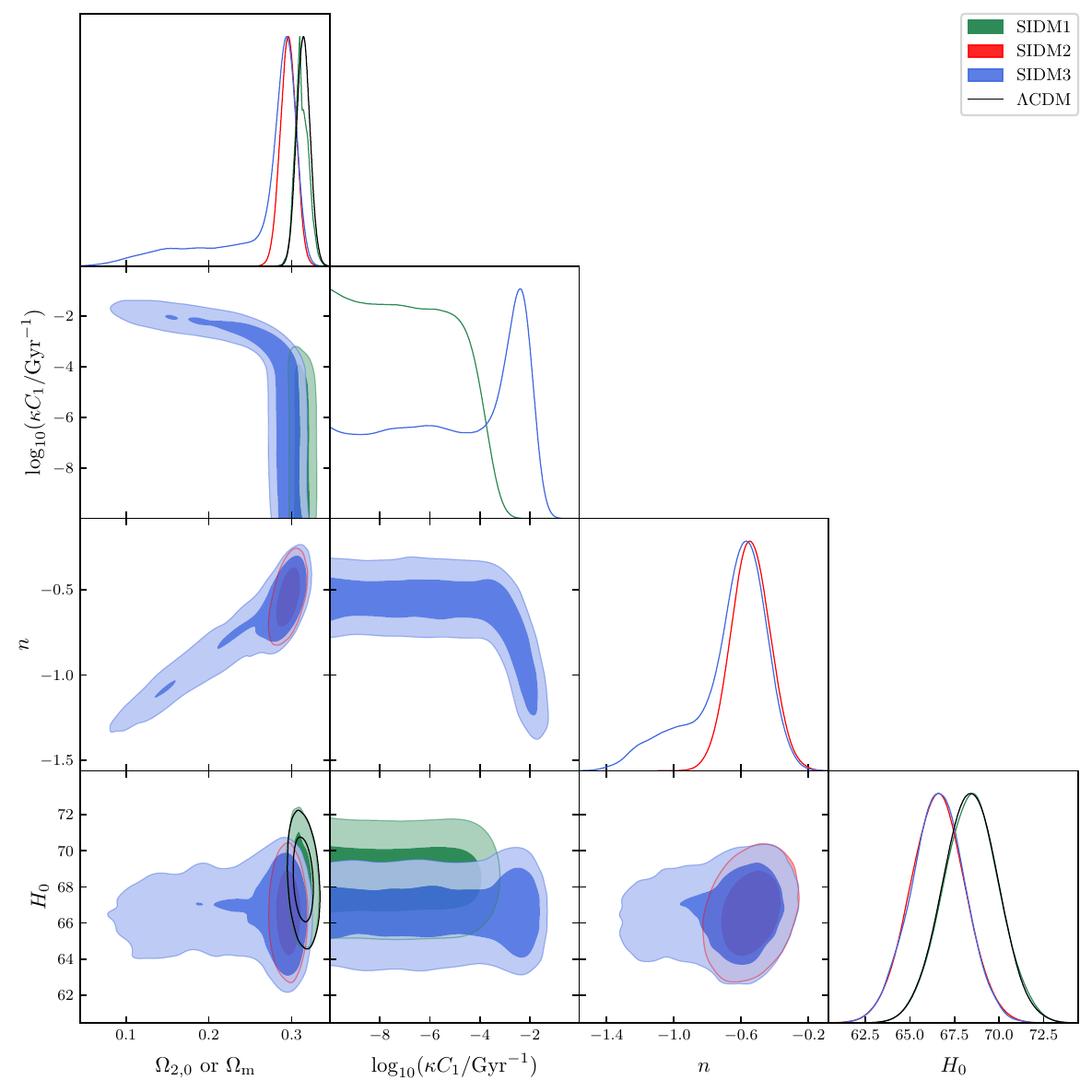}}
      \caption{
         2D marginalized contours and 1D posterior distributions at the $1\sigma$ and $2\sigma$ confidence levels for the cosmological parameters of the three SIDM models, derived from the combined DESI+CC+Pantheon+ dataset. Constraints on the $\Lambda$CDM model are shown for comparison.
      }
      \label{fig:5}
   \end{figure}

\subsection{Observational constraints}\label{sec:3.2}

   To demonstrate the constraining power of the latest observations, we employed three dataset combinations (DESI, DESI+Pantheon+, and DESI+DESY5) to place constraints on the cosmological parameters of the three SIDM models, assuming a spatially flat universe. The posterior 1D and 2D marginalized probability distributions, along with the 2D confidence contours for the cosmological parameters, are presented in Figs.~\ref{fig:2}--\ref{fig:4}. The marginalized best-fit parameter values and their $1\sigma$ uncertainties for all models and dataset combinations are listed in Table~\ref{tab:1}. For comparison, the corresponding $\Lambda$CDM constraints are also included in the same table.

   The constraint results for the SIDM1 model are shown in Fig.~\ref{fig:2}, which displays the 2D confidence contours (at $1\sigma$ and $2\sigma$) and the 1D marginalized posteriors for different dataset combinations. Our results show that the constraints obtained from the latest observations, including DESI BAOs, are significantly tighter than those previously reported in the literature. Specifically, the DESI BAO data alone yield well-determined constraints on all SIDM1 parameters at 68\% confidence level: $\Omega_{2,0} = 0.296\pm0.010$ and $\log_{10}(\kappa C_1)=-3.59$ (with an upper limit of $\log_{10}(\kappa C_1)=-3.20$). The effective annihilation term remains unconstrained toward smaller values. When combined with the Pantheon+ dataset, DESI BAO yields tighter constraints: $\Omega_{2,0} = 0.312\pm0.008$ and $\log_{10}(\kappa C_1)=-4.36$ (with an upper limit of $\log_{10}(\kappa C_1)=-3.93$). Including the DESY5 dataset instead shifts the matter density parameter to $\Omega_{2,0} = 0.314\pm0.008$ and tightens the constraint on the annihilation term, yielding a best-fit of $\log_{10}(\kappa C_1)=-4.29$ with an upper limit of $\log_{10}(\kappa C_1)=-3.89$. Across all dataset combinations, the current data consistently prefer an extremely small annihilation cross section in the SIDM1 scenario. This indicates that the BAO and SN Ia data, which are primarily sensitive to the expansion history, constrain the annihilation rate to an extremely small value, rendering the SIDM1 model practically indistinguishable from $\Lambda$CDM. The derived matter density parameter from DESI, DESI+Pantheon+, and DESI+DESY5 is in excellent agreement with the Planck 2018 TT,TE,EE+lowE+lensing result of $\Omega_{m} = 0.310\pm0.006$ \citep{2020A&A...641A...6P}. These findings differ from those of a previous study, which used a smaller SN Ia dataset and reported $\Omega_{2,0} = 0.271\pm0.032$ and $\log_{10}(\kappa C_1)=-4.42$ (with an upper limit of $\log_{10}(\kappa C_1)=-3.50$; \citealt{2011A&A...529A..61C}).

   The SIDM2 model behaves as a $w$CDM-like model through the introduction of one additional parameter, $n$. Figure~\ref{fig:3} presents the triangle plot of the constraint results for SIDM2 using different dataset combinations. With the DESI BAO data alone, we obtained $\Omega_{2,0} = 0.297\pm0.009$ and $n=-0.24^{+0.22}_{-0.24}$. The best-fit SIDM EoS is $w_{\rm{SIDM}}=-0.92^{+0.08}_{-0.07}$, which is consistent with the concordance $\Lambda$CDM value of $w=-1$ within $1\sigma$. Combining DESI BAO with Pantheon+ yielded more stringent constraints on all cosmological parameters: $\Omega_{2,0} =0.295 \pm 0.010$ and $n=-0.55 \pm 0.12$ ($w_{\rm{SIDM}}=-0.82 \pm 0.04$). For the DESI+DESY5 combination, we obtained $\Omega_{2,0} =0.297 \pm 0.009$ and $n=-0.48 \pm 0.12$ ($w_{\rm{SIDM}}=-0.84 \pm 0.04$). A strong degeneracy between $\Omega_{2,0}$ and $n$ is clearly visible in Fig.~\ref{fig:3}. A mild correlation between these parameters was also noted in a previous study using a smaller dataset \citep{2011A&A...529A..61C}, and our results confirm that this $\Omega_{2,0}$--$n$ degeneracy remains strong in the SIDM2 model. Furthermore, our results favor an effective quintessence-like model with $w_{\rm{SIDM}}>-1$, which is in tension with the findings of the earlier analysis \citep{2011A&A...529A..61C}.

   For the SIDM3 model, there are three free parameters: $\Omega_{2,0}$, $\kappa C_1$, and $n$ (equivalently, $w_{\rm{SIDM}}$). The constraint results from different dataset combinations are presented in Fig.~\ref{fig:4}, which reveals a strong degeneracy between the annihilation cross section and the DM EoS. Within this general SIDM framework, the constraints on the matter density parameter are $\Omega_{2,0}=0.289_{-0.005}^{+0.018}$, $0.243 \pm 0.070$, and $0.265_{-0.003}^{+0.051}$ for DESI, DESI+Pantheon+, and DESI+DESY5, respectively, all consistent with standard values from other astrophysical probes. The corresponding constraints on $\kappa C_1$ and $n$ (or $w_{\rm{SIDM}}$) are listed in Table~\ref{tab:1}. For the effective annihilation term, the estimated value in the SIDM3 model ranged from $\log_{10}(\kappa C_1)=-2.67$ (upper limit $-2.30$) to $\log_{10}(\kappa C_1)=-2.12$ (upper limit $-1.75$). The estimated values of $n=-0.27 \pm 0.27$ (DESI), $n=-0.74_{-0.14}^{+0.35}$ (DESI+Pantheon+), and $n=-0.61_{-0.11}^{+0.27}$ (DESI+DESY5) corresponded to SIDM EoS values of $w_{\rm{SIDM}}=-0.91 \pm 0.09$, $w_{\rm{SIDM}}=-0.75_{-0.12}^{+0.05}$, and $w_{\rm{SIDM}}=-0.80_{-0.09}^{+0.04}$, respectively. All of these constraints are most consistent with an effective quintessence-like dark energy model, representing the most robust conclusion drawn from the current BAO+SN Ia datasets.

   Finally, we performed a joint analysis of the three SIDM models using the combined DESI+CC+Pantheon+ dataset. The $1\sigma$ and $2\sigma$ confidence contours for the parameter estimates are shown in Fig.~\ref{fig:5}, and the marginalized best-fit parameters with $1\sigma$ uncertainties are summarized in Table~\ref{tab:1}. The joint analysis of standard rulers, standard candles, and standard clocks yielded the following constraints: for SIDM1, $\log_{10}(\kappa C_1)=-4.32$ (upper limit $-3.90$); for SIDM2, $n=-0.54 \pm 0.12$ ($w_{\rm{SIDM}}=-0.82 \pm 0.04$); and for SIDM3, $\log_{10}(\kappa C_1)=-2.24$ (upper limit $-1.87$) and $n=-0.67_{-0.09}^{+0.26}$ ($w_{\rm{SIDM}}=-0.78_{-0.09}^{+0.03}$). The corresponding $H_0$ determinations are $H_0=68.4 \pm 1.6~{\rm km\,s^{-1}\,Mpc^{-1}}$ for SIDM1, $H_0=66.6 \pm 1.6~{\rm km\,s^{-1}\,Mpc^{-1}}$ for SIDM2, and $H_0=66.6 \pm 1.5~{\rm km\,s^{-1}\,Mpc^{-1}}$ for SIDM3. Across all three SIDM models, the inferred Hubble constant is in agreement with the Planck 2018 result ($67.4\pm0.5~{\rm km\,s^{-1}\,Mpc^{-1}}$; \citealt{2020A&A...641A...6P}), but in tension with the local distance ladder measurement from SH0ES ($73.04\pm1.04~{\rm km\,s^{-1}\,Mpc^{-1}}$; \citealt{2022ApJ...934L...7R}). However, one should note that these values are smaller than the \(\Lambda\)CDM Planck value when considering the central estimates for SIDM2 and SIDM3. Compared with the standard \(\Lambda\)CDM cosmology, these two SIDM scenarios do not shift \(H_0\) toward the local distance-ladder value, which is opposite to what is required to alleviate the Hubble tension.

   \begin{table}
      \caption{Comparison of the SIDM models against the $\Lambda$CDM model.}
      \label{tab:2}
      \centering
      \begin{tabular}{l|ccc}
         \hline\hline
         Model/Dataset & $\Delta\chi^2$ & Significance & $\ln\mathcal{B}_{ij}$ \\
         \hline
         SIDM1 \\
         DESI & $-0.34$ & $0.15\sigma$ & $0.80$ \\
         DESI+Pantheon+ & $-2.20$ & $1.1\sigma$ & $1.27$ \\
         DESI+DESY5 & $-0.36$ & $0.12\sigma$ & $0.59$ \\
         DESI+CC+Pantheon+ & $-2.06$ & $1.0\sigma$ & $1.15$ \\
         \hline
         SIDM2 \\
         DESI & $-1.08$ & $0.53\sigma$ & $-3.09$ \\
         DESI+Pantheon+ & $-22.04$ & $4.6\sigma$ & $6.69$ \\
         DESI+DESY5 & $-17.25$ & $4.0\sigma$ & $4.29$ \\
         DESI+CC+Pantheon+ & $-21.71$ & $4.5\sigma$ & $5.57$ \\
         \hline
         SIDM3 \\
         DESI & $-3.22$ & $0.84\sigma$ & $-1.80$\\
         DESI+Pantheon+ & $-29.56$ & $4.9\sigma$ & $10.15$ \\
         DESI+DESY5 & $-22.45$ & $4.2\sigma$ & $6.94$ \\
         DESI+CC+Pantheon+ & $-28.47$ & $4.8\sigma$ & $8.72$ \\
         \hline
      \end{tabular}
      \tablefoot{Differences in $\chi^2$ and log-Bayes factors ($\ln\mathcal{B}_{ij}$) between the SIDM models and the reference $\Lambda$CDM model. The significance is computed from $\Delta\chi^2$ accounting for the extra degrees of freedom in each SIDM model.}
   \end{table}

   \begin{figure}
      \resizebox{\hsize}{!}{\includegraphics{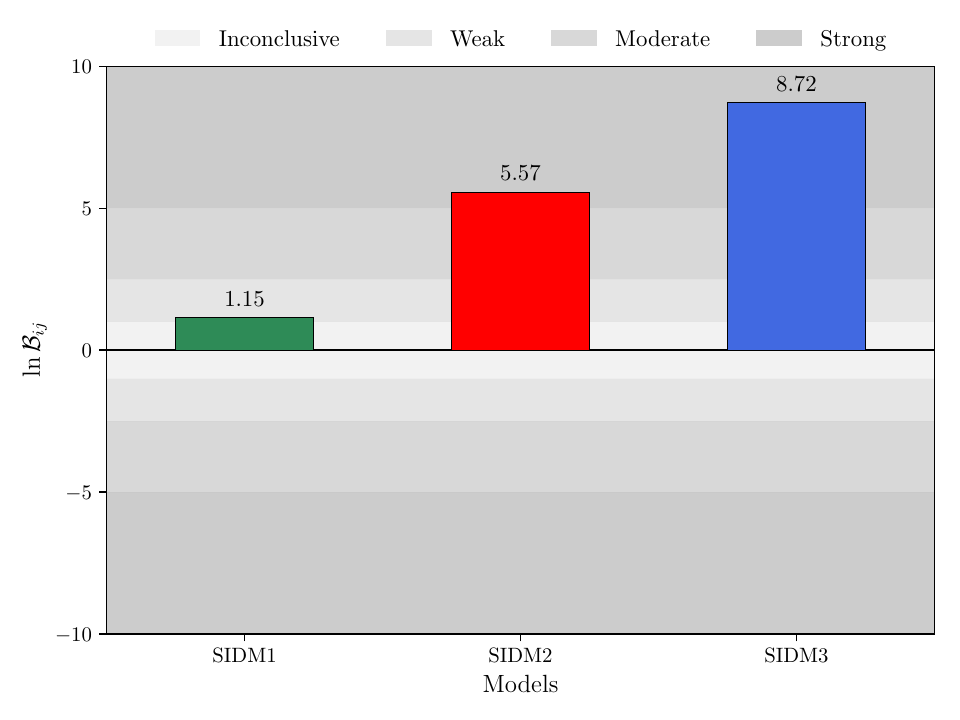}}
      \caption{
         Comparison of the Bayesian evidence for the SIDM models and the $\Lambda$CDM model, based on the combined DESI+CC+Pantheon+ dataset. The log-Bayes factor ($\ln\mathcal{B}_{ij}$) for the SIDM model relative to $\Lambda$CDM and its associated strength of evidence are used to assess the performance of different models. Note that a positive value indicates the preference for the SIDM model.
      }
      \label{fig:6}
   \end{figure}

\subsection{Statistical analysis}\label{sec:3.3}

The statistical analysis is essential to diagnose different models. Hence, we applied the Bayesian evidence to compare these SIDM models and discussed how strongly they are favored by the observational datasets.

We used the Bayesian evidence (marginal likelihood) to assess whether the SIDM models are preferred over the $\Lambda$CDM model given the current observational data. The Bayesian evidence $\mathcal{Z}$ is defined as 
   \begin{equation}
      \mathcal{Z} = \int \mathcal{L}(\theta)\,\pi(\theta)\,\mathrm{d}\theta,
   \end{equation}
   where $\mathcal{L}(\theta)$ is the likelihood and $\pi(\theta)$ is the prior. We computed $\ln\mathcal{Z}$ using the $k$th-nearest-neighbor estimator of \citet{2017arXiv170403472H}. The key idea is to estimate the posterior density at each MCMC sample from the distance to its $k$th nearest neighbor, and then to use the identity $\mathcal{Z} = \tilde p(\theta) / p(\theta)$ (where $\tilde p(\theta) = \mathcal{L}(\theta)\pi(\theta)$ is the un-normalized posterior) to obtain the evidence. The Bayes factor $\ln\mathcal{B}_{ij} = \ln\mathcal{Z}_i - \ln\mathcal{Z}_j$ quantifies the evidence in favor of model $i$ over model $j$. Following \citet{Kass01061995}, we interpret $1<|\ln\mathcal{B}_{ij}|<2.5$ as weak evidence, $2.5<|\ln\mathcal{B}_{ij}|<5$ as moderate evidence, and $|\ln\mathcal{B}_{ij}|>5$ as strong evidence in favor of the preferred model. In Table~\ref{tab:2}, we listed the values of $\Delta\chi^2 = \chi_{\rm{SIDM}}^2 - \chi_{\Lambda\rm{CDM}}^2$, the statistical significance in terms of $\sigma$ after accounting for the extra degrees of freedom $k$, and $\ln\mathcal{B}_{ij}$ for the different SIDM models across independent and joint analyses (DESI, DESI+Pantheon+/DESY5, and DESI+CC+Pantheon+). Here, $i$ and $j$ denote the SIDM and $\Lambda$CDM model, respectively. Focusing on the dataset combinations DESI, DESI+Pantheon+, and DESI+DESY5, the three SIDM models can be divided into two classes based on the number of free parameters: two-parameter models (SIDM1 and SIDM2) and a three-parameter model (SIDM3). Among all models, SIDM3 has the largest Bayes factor, indicating strong evidence over $\Lambda$CDM, and SIDM2 is also strongly favored. The SIDM1 model yields $\ln\mathcal{B}_{ij} \simeq 1$. A graphical summary of the $\ln\mathcal{B}_{ij}$ values for the DESI+CC+Pantheon+ dataset is provided in Fig.~\ref{fig:6}. The $\ln\mathcal{B}_{ij}$ values relative to $\Lambda$CDM are: $1.15$ for SIDM1, $5.57$ for SIDM2, and $8.72$ for SIDM3. These results confirm that SIDM2 and SIDM3 provide significantly better fits to the observational data, strongly preferred over the concordance standard $\Lambda$CDM model.

   \begin{figure*}
      \sidecaption
      \includegraphics[width=12cm]{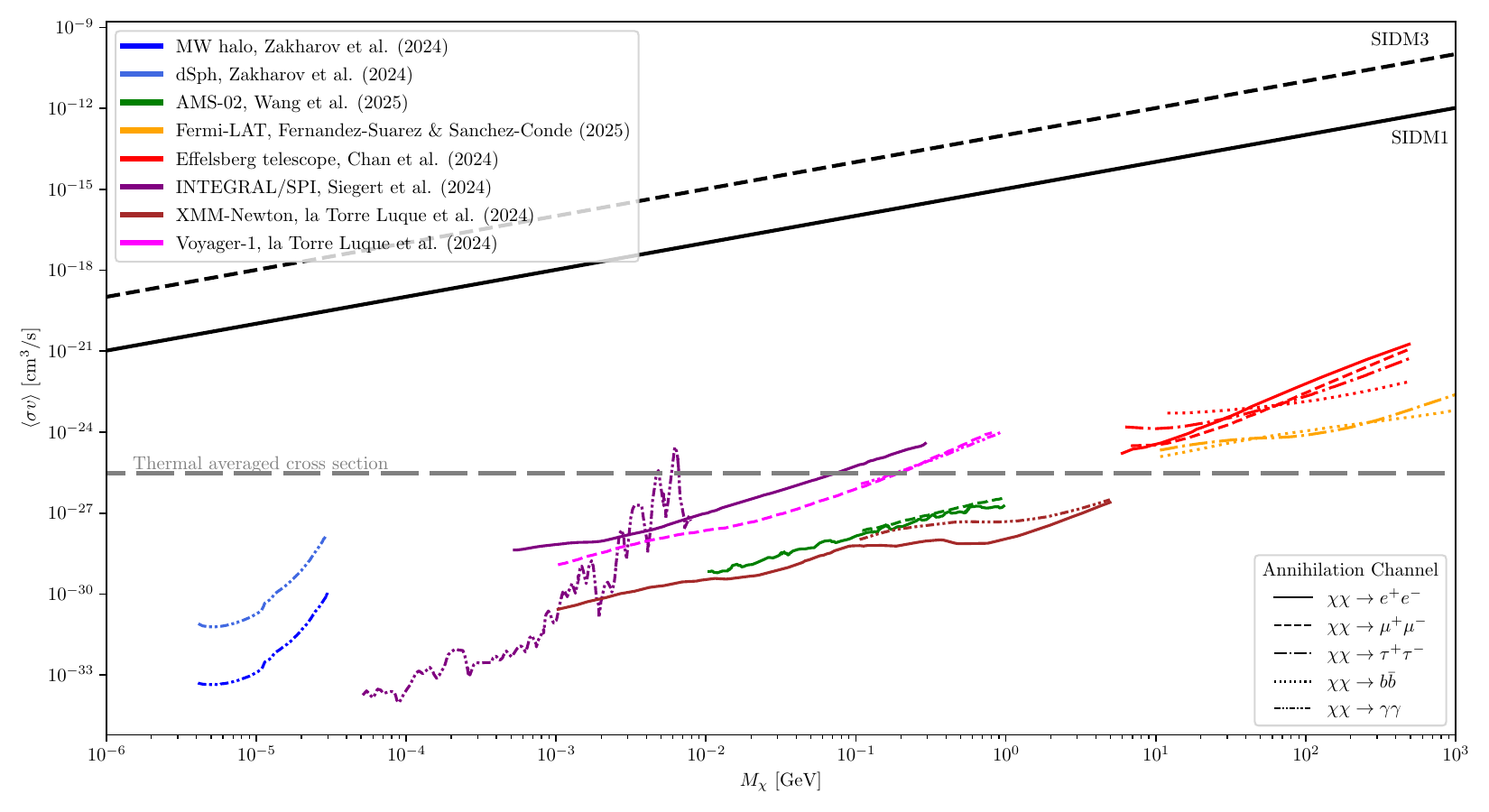}
      \caption{Velocity-averaged DM annihilation cross section, $\langle\sigma v\rangle$, as a function of the DM particle mass, $M_{\chi}$, within the SIDM framework. Different colors represent different datasets, and different line styles denote different annihilation channels. The dashed gray line marks the thermal relic abundance cross section $\langle\sigma v \rangle \simeq 3\times 10^{-26}$~cm${}^3$/s. The solid and dashed black lines indicate the upper limits on $\langle\sigma v\rangle$ for SIDM1 and SIDM3, respectively, derived from the DESI+CC+Pantheon+ dataset combination.
      }
      \label{fig:7}
   \end{figure*}
         
   \begin{table*}
      \caption{Observational constraints on the self-interaction cross section per unit mass ($\sigma/M_{\chi}$) from this and previous analyses.}
      \label{tab:3}
      \centering
      \begin{tabular}{l|ccc}
         \hline\hline
         Dataset & $\sigma/M_{\chi}$ & $v_{\rm{rel}}$ & References \\
         \hline
         Previous analysis \\
         Halo shapes/ellipticity & $\lesssim 1~\rm cm^2\,g^{-1}$ & $1300~{\rm km\,s^{-1}}$ & 1 \\
         Substructure mergers & $\lesssim 1~\rm cm^2\,g^{-1}$ & $\sim 500 - 4000~{\rm km\,s^{-1}}$ & 2, 3 \\
         Bullet Cluster & $\lesssim 0.7~\rm cm^2\,g^{-1}$ & $4000~{\rm km\,s^{-1}}$ & 4 \\
         \hline
         This analysis (SIDM1) \\
         Low-redshift dataset combination & $\lesssim 6~\rm cm^2\,g^{-1}$ & $\sim 1000~{\rm km\,s^{-1}}$ & 5 \\
         CMB & $\lesssim 0.2~\rm cm^2\,g^{-1}$ &  $\sim 1000~{\rm km\,s^{-1}}$ & 5 \\
         DESI+CMB & $\lesssim 0.01~\rm cm^2\,g^{-1}$ &  $\sim 1000~{\rm km\,s^{-1}}$ & 5 \\
         \hline
      \end{tabular}
      \tablebib{
         1. \citet{2013MNRAS.430..105P}; 2. \citet{2015IJMPD..2450071C}; 3. \citet{2018ApJ...869..104W}; 4. \citet{2008ApJ...679.1173R}; 5. This work.
      }
   \end{table*}

\subsection{Dark matter annihilation cross section}\label{sec:3.4}

   The cosmological constraints on the effective annihilation parameter $\kappa$ derived above can be directly translated into upper limits on the velocity-averaged annihilation cross section $\langle\sigma v\rangle$ and the DM particle mass $M_{\chi}$. This mapping enables a direct comparison between our large-scale phenomenological constraints and the extensive set of bounds obtained from indirect detection searches at galactic and sub-galactic scales, which typically assume specific annihilation channels within a weakly interacting massive particle framework. Indirect detection analyses have placed stringent upper bounds on $\langle\sigma v\rangle$ over a wide range of $M_{\chi}$ and for numerous annihilation final states, including $e^+e^-$, $\mu^+\mu^-$, $\tau^+\tau^-$, $b\bar{b}$, and $\gamma\gamma$. For example, \citet{2024PhRvD.110l3026Z} derived constraints using SRG/ART-XC observations of the Milky Way halo and 33 Local Group dwarf spheroidal (dSph) galaxies, covering DM masses down to $10^{-5}$ GeV. \citet{2025EPJC...85.1348W} analyzed Alpha Magnetic Spectrometer (AMS-02) cosmic-ray data for masses ranging from $10^{-2}$ GeV to 1~GeV. \citet{2025JCAP...09..003F} used \textit{Fermi} Large Area Telescope (LAT) data to search for DM annihilation in 11 stellar streams, probing masses from 10~GeV to $10^4$ GeV. \citet{2024ApJ...962..141C} obtained constraints from archival Effelsberg radio observations of M33 up to 500~GeV. \citet{2024MNRAS.528.3433S} employed 16 years of INTEGRAL/SPI (the Spectrometer on INTEGRAL) $\gamma$-ray data to search for light DM in the Milky Way halo, with sensitivity to masses around $10^{-3}$ GeV. \citet{2024ApJ...968...46D} analyzed \textit{XMM-Newton} and Voyager-1 cosmic-ray data, covering masses from 1~MeV to several GeV. The constraints from these indirect detection probes are presented in Fig.~\ref{fig:7}. 

   The combined DESI+CC+Pantheon+ dataset yields $68\%$ upper limits on the effective annihilation parameter of $\kappa \lesssim 10^{-15}~{\rm cm^3\,s^{-1}\,GeV^{-1}}$ for SIDM1 and $\kappa \lesssim 10^{-13}~{\rm cm^3\,s^{-1}\,GeV^{-1}}$ for SIDM3 (see Fig.~\ref{fig:7}). By contrast, various indirect detection probes imply $\kappa \sim 10^{-25}~{\rm cm^3\,s^{-1}\,GeV^{-1}}$, which is approximately 10--12 orders of magnitude smaller than the values favored by our SIDM scenario. This large discrepancy indicates that the SIDM models considered here require significantly stronger self-scattering interactions than conventional weakly interacting massive particle candidates. Adopting the characteristic velocity of $\sim 1000~{\rm km\,s^{-1}}$ for DM on large cosmological scales \citep{2018PhR...730....1T}, the present DESI+CC+Pantheon+ constraints imply $\sigma/M_{\chi} \approx 5.8~{\rm cm^2\,g^{-1}}$ for SIDM1 and $\approx 580~{\rm cm^2\,g^{-1}}$ for SIDM3, both of which are substantially larger than the competitive bound of $\sigma/M_{\chi} \lesssim 0.1~{\rm cm^2\,g^{-1}}$. Nevertheless, theoretical arguments demonstrate that the quantity $\kappa C_1$ constrained from low-redshift data ($z \lesssim 2$) cannot adequately describe the early Universe at $z \sim 1000$. To supplement the low-redshift constraints, we additionally adopted the CMB distance priors $\{l_a, R, z_*\}$ from Planck 2018 \citep{2020A&A...641A...6P, 2019JCAP...02..028C} to impose high-redshift constraints. Since these priors are derived under the $\Lambda$CDM assumption, we applied them exclusively to the SIDM1 model, which formally reduces to $\Lambda$CDM in the limit $\kappa \to 0$. The CMB dataset alone yields $\log_{10}(\kappa C_1/{\rm Gyr}^{-1}) < -5.25$ for SIDM1, corresponding to $\sigma/M_{\chi} \lesssim 0.2~{\rm cm^2\,g^{-1}}$ at $v_{\rm{rel}} \simeq 1000~{\rm km\,s^{-1}}$, consistent with constraints from galaxy clusters \citep{2016PhRvL.116d1302K, 2018ApJ...853..109E}. The combined DESI+CMB dataset further tightens this to $\log_{10}(\kappa C_1/{\rm Gyr}^{-1}) < -6.85$, corresponding to $\sigma/M_{\chi} \lesssim 0.01~{\rm cm^2\,g^{-1}}$. These results, together with comparisons to the recent literature, are summarized in Table~\ref{tab:3}. 
   
   Therefore, our constraints within the SIDM framework clearly indicate that $\sigma/M_{\chi}$ is velocity-dependent, with a preferred cosmological scaling of $\sigma/M_{\chi} \propto 1/v_{\rm{rel}}$. Standard SIDM models based on a Yukawa potential naturally predict velocity-dependent self-interaction cross sections \citep{2018PhR...730....1T, 2013PhRvD..87k5007T}, with scalings close to $\sigma/M_{\chi} \propto 1/v_{\rm{rel}}^4$ at cluster scales and nearly constant behavior at galaxy scales. Resonant self-interactions provide an attractive realization of such a velocity dependence without the need for a light mediator \citep{2019PhRvL.122g1103C}. The power-law index of the velocity dependence is highly sensitive to both the DM mass and the mediator mass, whereas our phenomenological SIDM description does not incorporate detailed particle-physics properties. Future theoretical work is therefore needed to develop a consistent interpretation of the scale-dependent evolution of DM annihilation and self-interaction cross sections from cluster to cosmological scales.

\section{Conclusions}\label{sec:4}

In this paper, we have revisited the SIDM scenario, which we originally proposed to resolve the small-scale problems of the standard CDM paradigm and additionally provides a mechanism for driving late-time cosmic acceleration. Using the latest DESI DR2 BAO data, SN Ia data from Pantheon+ and DESY5, and Hubble parameter measurements from CCs, we have placed tight constraints on the effective annihilation cross section and the EoS of DM across three representative SIDM models. Our key findings are summarized as follows. 

For the SIDM1 model, which focuses on DM particle annihilation, the joint analysis of DESI DR2, Pantheon+, and CCs places a stringent upper limit on the effective annihilation cross section of $\log_{10}(\kappa C_1/{\rm Gyr}^{-1}) < -3.90$ (best-fit value $-4.32$). This result, consistent with an extremely small annihilation rate, renders SIDM1 practically indistinguishable from $\Lambda$CDM cosmology and differs from the findings of earlier work \citep{2011A&A...529A..61C}. The derived matter density parameter is fully consistent with the Planck 2018 results \citep{2020A&A...641A...6P}. For the SIDM2 model, the SIDM fluid effectively acts as a dynamical dark energy component, with the combined datasets yielding a mildly dynamical EoS of $w_{\rm{SIDM}}=-0.82 \pm 0.04$, favoring an effective quintessence-like model with $w_{\rm{SIDM}} > -1$. For the more general SIDM3 model, we find a strong degeneracy between the annihilation cross section and the EoS, with $w_{\rm{SIDM}}=-0.78_{-0.09}^{+0.03}$ and $\log_{10}(\kappa C_1/{\rm Gyr}^{-1})<-1.87$. Across all three SIDM models, the inferred Hubble constant is in excellent agreement with the Planck 2018 result \citep{2020A&A...641A...6P}, while exhibiting tension with the local distance-ladder measurement from SH0ES \citep{2022ApJ...934L...7R}. Based on the Bayesian evidence, the SIDM2 and SIDM3 models are strongly preferred over the standard $\Lambda$CDM cosmology, with log-Bayes factors $\ln\mathcal{B}_{ij} = 5.57$ and $8.72$, while SIDM1 is weakly preferred over $\Lambda$CDM ($\ln\mathcal{B}_{ij} = 1.15$) due to its highly suppressed annihilation term. Furthermore, the cosmological constraints on the effective annihilation parameter can be translated into bounds on the velocity-averaged annihilation cross section and the self-interaction cross section per unit mass. While low-redshift data alone imply a relatively large cross section, the inclusion of CMB distance priors significantly tightens the constraint at cluster-scale velocities, highlighting the necessity of a velocity-dependent cross section to reconcile cosmological and astrophysical observations.

There are several remarks that remain to be clarified as follows. Firstly, since the effective EoS of SIDM model $w_{\rm SIDM}$ is constant, it does not evolve with redshift. For the combined DESI+CC+Pantheon+ dataset, the best-fit values $w_{\rm SIDM}=-0.82\pm0.04$ in SIDM2 and $w_{\rm SIDM}=-0.78_{-0.09}^{+0.03}$ in SIDM3 place the SIDM fluid in the quintessence regime $w_{\rm SIDM}>-1$, where the null and dominant energy conditions, $\rho+P\ge0$ and $\rho\ge|P|$, are satisfied while the strong energy condition is violated, $\rho+3P<0$, as required for the accelerated expansion discussed above. Because $w_{\rm SIDM}$ remains above $-1$, the model does not lead to phantom-like pathologies such as a future big-rip singularity. Unlike single-field scalar models, where crossing $w=-1$ typically involves instabilities \citep{2003PhRvL..91g1301C} and thermodynamic arguments disfavor $w<-1$ \citep{2008PhLB..669..266P}, the SIDM effective pressure originates from particle creation and annihilation processes rather than from a scalar Lagrangian, and therefore does not suffer from the same pathology. The quintessence-like EoS should thus be interpreted as a phenomenological description of the background dynamics rather than as evidence of a fundamental quintessence field. If the SIDM component were formally treated as a perfect barotropic fluid with $P=w_{\rm SIDM}\rho$, the adiabatic sound speed would be $c_{\rm s}^{2}=\mathrm{d}P/\mathrm{d}\rho=w_{\rm SIDM}<0$, which would signal a small-scale gradient instability. This apparent pathology does not apply to the SIDM framework, because the fluid is non-adiabatic, with entropy generated by the creation and annihilation processes. The correct treatment, based on the collisional Boltzmann equation and its neo-Newtonian approximation, shows that the linear growth of density fluctuations is stable and recovers the $\Lambda$CDM growth history \citep{2011PhRvD..84f3511J, 2014JCAP...10..042L}. The apparent gradient instability is therefore an artifact of the barotropic-fluid approximation rather than a genuine pathology of the model. We note that the present analysis is restricted to the background expansion, and the growth of linear density perturbations within the SIDM scenarios has not yet been tested against growth-rate data -- an analysis that the original interacting DM study \citep{2009A&A...507...47B} explicitly left for future work. Nevertheless, for the SIDM2 model, whose effective EoS is constant, the growth of a pressureless matter perturbation in the presence of a smooth effective dark-energy component follows the standard growth equation, and the growth rate can be approximated by $f(a)\simeq\Omega_{2,0}(a)^{\gamma}$ with the growth index $\gamma\simeq3(w_{\rm SIDM}-1)/(6w_{\rm SIDM}-5)\simeq0.55$ for the best-fit parameters of Table~\ref{tab:1}, essentially indistinguishable from the $\Lambda$CDM value $\gamma\simeq0.545$ \citep{1998ApJ...508..483W, 2005PhRvD..72d3529L}. The predicted $f\sigma_8(z)$ is therefore degenerate with $\Lambda$CDM within the current redshift-space-distortion data \citep{2017PhRvD..96b3542N}, so the growth of structure does not yet discriminate between the SIDM2 scenario and the concordance model. For the SIDM1 and SIDM3 scenarios ($\kappa\neq0$), a dedicated perturbation analysis is left for future work. Secondly, in the limit $\kappa\to0$, the SIDM2 model is mathematically equivalent to the gravitational matter creation model of \citet{2008CQGra..25t5006L}, whose physical origin dates back to the gravitational particle-production mechanism proposed by \citet{1970JETPL..12..307Z}. This equivalence connects the phenomenological SIDM approach to a framework with a foundation in quantum field theory in curved space-time, which has been shown to satisfy the second law of thermodynamics and to approach thermodynamic equilibrium at late times \citep{2008CQGra..25t5006L, 2014JCAP...10..042L}. The viability of this particle-creation mechanism has been further verified at the perturbative level \citep{2011PhRvD..84f3511J}. The SIDM2 constraints presented in this work can therefore be interpreted as observational bounds on the gravitational particle creation rate, complementing earlier analyses of matter-creation cosmologies \citep{2010PhRvD..82b3504B, 2012PhRvD..86j3534L, 2016MNRAS.459..673N}. A limitation of the SIDM1 model ($n=0$, $\kappa\neq0$) lies in the fact that the exact analytical solution for the Hubble parameter exhibits a singularity at a maximum redshift $z_{\max}$, beyond which $H(z)$ becomes imaginary. This singularity arises because the exponential damping factor $e^{-2\kappa C_1 t}$ in Eq.~(\ref{eq:8}) eventually dominates the $a^{-3}$ term, causing the energy density to vanish at a finite scale factor. For typical values of $\kappa C_1$ constrained by low-redshift data, $z_{\max}\sim\mathcal{O}(1-10)$, which is well below the recombination epoch and prevents the SIDM1 expansion history from being extrapolated to early times. This limitation does not affect SIDM2, whose $\kappa=0$ limit bypasses the singularity, but it still influences SIDM3, for which no analytical form exists. Finally, it should be noted that the adopted functional form of the creation rate $\Psi(a)$ has no fundamental quantum field theory derivation and is selected primarily by the requirement of analytic solvability of the resulting Riccati equation (see Sect.~\ref{sec:2}). Consequently, the parameters $n$ and $\kappa C_1$ must be tuned carefully to satisfy simultaneously the cosmic-acceleration threshold and the expansion data, a fine-tuning concern analogous to, although distinct from, the usual cosmological-constant problem. Moreover, the SIDM framework does not provide a natural attractor mechanism that selects the observed ratio of matter to effective dark-energy densities at the present epoch, so the coincidence problem remains as open an issue as in $\Lambda$CDM or quintessence. We emphasize that the SIDM framework investigated here should be regarded as a phenomenological reconstruction of the background expansion history rather than a complete physical model. Although the SIDM2 and SIDM3 scenarios are statistically preferred over $\Lambda$CDM, this improved goodness-of-fit does not translate into a viable physical mechanism to solve the \(H_0\) tension. On the contrary, the direction of the shift in \(H_0\) indicates that the underlying physics of these models exacerbates the discrepancy with the local measurement.

Summarizing, the present paper is only the first step toward elaborating the scheme to the potential of SIDM as a unified framework capable of simultaneously addressing small-scale structure problems and providing an effective dark energy component that drives cosmic acceleration. While this analysis still does not fully explore all physical SIDM scenarios, our results highlight the importance of including future observations from \textit{Euclid} \citep{2025A&A...697A...1E}, the Legacy Survey of Space and Time (LSST) at the \textit{Vera C. Rubin} Observatory \citep{2019ApJ...873..111I}, and the Simons Observatory \citep{2019JCAP...02..056A}, which will further improve constraints on SIDM parameters and help elucidate the physical nature of DM self-interactions \citep{2020PhRvL.125k1105Y,PhysRevLett.116.041302}.

\begin{acknowledgements}

This work is supported by National Natural Science Foundation of China (Nos. 12675058; 12433001); Beijing Natural Science Foundation (No. 1242021); and the Fundamental Research Funds for Central Universities.

\end{acknowledgements}

\bibliographystyle{aa}
\bibliography{reference}

@article{1970JETPL..12..307Z,
  author  = {{Zel'dovich}, Ya. B.},
  title   = {{Particle Production in Cosmology}},
  journal = {Soviet Journal of Experimental and Theoretical Physics Letters},
  year    = 1970,
  month   = jan,
  volume  = {12},
  pages   = {307},
  adsurl  = {https://ui.adsabs.harvard.edu/abs/1970JETPL..12..307Z}
}

@article{1988PhRvD..37.3406R,
  author   = {{Ratra}, Bharat and {Peebles}, P.~J.~E.},
  title    = {{Cosmological consequences of a rolling homogeneous scalar field}},
  journal  = {\prd},
  year     = 1988,
  month    = jun,
  volume   = {37},
  number   = {12},
  pages    = {3406-3427},
  doi      = {10.1103/PhysRevD.37.3406},
  adsurl   = {https://ui.adsabs.harvard.edu/abs/1988PhRvD..37.3406R}
}

@article{1988PNAS...85.7428P,
  author  = {{Prigogine}, I. and {Geheniau}, J. and {Gunzig}, E. and {Nardone}, P.},
  title   = {{Thermodynamics of Cosmological Matter Creation}},
  journal = {Proceedings of the National Academy of Science},
  year    = 1988,
  month   = oct,
  volume  = {85},
  number  = {20},
  pages   = {7428-7432},
  doi     = {10.1073/pnas.85.20.7428},
  adsurl  = {https://ui.adsabs.harvard.edu/abs/1988PNAS...85.7428P}
}

@article{1989GReGr..21..767P,
  author  = {{Prigogine}, I. and {Geheniau}, J. and {Gunzig}, E. and {Nardone}, P.},
  title   = {{Thermodynamics and cosmology}},
  journal = {General Relativity and Gravitation},
  year    = 1989,
  month   = aug,
  volume  = {21},
  number  = {8},
  pages   = {767-776},
  doi     = {10.1007/BF00758981},
  adsurl  = {https://ui.adsabs.harvard.edu/abs/1989GReGr..21..767P}
}

@article{1992PhLA..162..223C,
  author  = {{Calv{\~a}o}, M.~O. and {Lima}, J.~A.~S. and {Waga}, I.},
  title   = {{On the thermodynamics of matter creation in cosmology}},
  journal = {Physics Letters A},
  year    = 1992,
  month   = feb,
  volume  = {162},
  number  = {3},
  pages   = {223-226},
  doi     = {10.1016/0375-9601(92)90437-Q},
  adsurl  = {https://ui.adsabs.harvard.edu/abs/1992PhLA..162..223C}
}

@article{1998AJ....116.1009R,
  author        = {{Riess}, Adam G. and {Filippenko}, Alexei V. and {Challis}, Peter and {Clocchiatti}, Alejandro and {Diercks}, Alan and {Garnavich}, Peter M. and {Gilliland}, Ron L. and {Hogan}, Craig J. and {Jha}, Saurabh and {Kirshner}, Robert P. and {Leibundgut}, B. and {Phillips}, M.~M. and {Reiss}, David and {Schmidt}, Brian P. and {Schommer}, Robert A. and {Smith}, R. Chris and {Spyromilio}, J. and {Stubbs}, Christopher and {Suntzeff}, Nicholas B. and {Tonry}, John},
  title         = {{Observational Evidence from Supernovae for an Accelerating Universe and a Cosmological Constant}},
  journal       = {\aj},
  year          = 1998,
  month         = sep,
  volume        = {116},
  number        = {3},
  pages         = {1009-1038},
  doi           = {10.1086/300499},
  archiveprefix = {arXiv},
  eprint        = {astro-ph/9805201},
  primaryclass  = {astro-ph},
  adsurl        = {https://ui.adsabs.harvard.edu/abs/1998AJ....116.1009R}
}

@article{1998ApJ...508..483W,
  author        = {{Wang}, Limin and {Steinhardt}, Paul J.},
  title         = {{Cluster abundance constraints for cosmological models with a time-varying, spatially inhomogeneous energy component with negative pressure}},
  journal       = {\apj},
  year          = 1998,
  month         = dec,
  volume        = {508},
  number        = {2},
  pages         = {483-490},
  doi           = {10.1086/306436},
  archiveprefix = {arXiv},
  eprint        = {astro-ph/9804015},
  primaryclass  = {astro-ph},
  adsurl        = {https://ui.adsabs.harvard.edu/abs/1998ApJ...508..483W}
}

@article{1998PhRvL..80.1582C,
  author        = {{Caldwell}, R.~R. and {Dave}, Rahul and {Steinhardt}, Paul J.},
  title         = {{Cosmological Imprint of an Energy Component with General Equation of State}},
  journal       = {\prl},
  year          = 1998,
  month         = feb,
  volume        = {80},
  number        = {8},
  pages         = {1582-1585},
  doi           = {10.1103/PhysRevLett.80.1582},
  archiveprefix = {arXiv},
  eprint        = {astro-ph/9708069},
  primaryclass  = {astro-ph},
  adsurl        = {https://ui.adsabs.harvard.edu/abs/1998PhRvL..80.1582C}
}

@article{1999ApJ...517..565P,
  author        = {{Perlmutter}, S. and {Aldering}, G. and {Goldhaber}, G. and {Knop}, R.~A. and {Nugent}, P. and {Castro}, P.~G. and {Deustua}, S. and {Fabbro}, S. and {Goobar}, A. and {Groom}, D.~E. and {Hook}, I.~M. and {Kim}, A.~G. and {Kim}, M.~Y. and {Lee}, J.~C. and {Nunes}, N.~J. and {Pain}, R. and {Pennypacker}, C.~R. and {Quimby}, R. and {Lidman}, C. and {Ellis}, R.~S. and {Irwin}, M. and {McMahon}, R.~G. and {Ruiz-Lapuente}, P. and {Walton}, N. and {Schaefer}, B. and {Boyle}, B.~J. and {Filippenko}, A.~V. and {Matheson}, T. and {Fruchter}, A.~S. and {Panagia}, N. and {Newberg}, H.~J.~M. and {Couch}, W.~J. and {Project}, The Supernova Cosmology},
  title         = {{Measurements of {\ensuremath{\Omega}} and {\ensuremath{\Lambda}} from 42 High-Redshift Supernovae}},
  journal       = {\apj},
  year          = 1999,
  month         = jun,
  volume        = {517},
  number        = {2},
  pages         = {565-586},
  doi           = {10.1086/307221},
  archiveprefix = {arXiv},
  eprint        = {astro-ph/9812133},
  primaryclass  = {astro-ph},
  adsurl        = {https://ui.adsabs.harvard.edu/abs/1999ApJ...517..565P}
}

@article{1999ApJ...522...82K,
  author        = {{Klypin}, Anatoly and {Kravtsov}, Andrey V. and {Valenzuela}, Octavio and {Prada}, Francisco},
  title         = {{Where Are the Missing Galactic Satellites?}},
  journal       = {\apj},
  year          = 1999,
  month         = sep,
  volume        = {522},
  number        = {1},
  pages         = {82-92},
  doi           = {10.1086/307643},
  archiveprefix = {arXiv},
  eprint        = {astro-ph/9901240},
  primaryclass  = {astro-ph},
  adsurl        = {https://ui.adsabs.harvard.edu/abs/1999ApJ...522...82K}
}

@article{2000PhRvL..84.3760S,
  author        = {{Spergel}, David N. and {Steinhardt}, Paul J.},
  title         = {{Observational Evidence for Self-Interacting Cold Dark Matter}},
  journal       = {\prl},
  year          = 2000,
  month         = apr,
  volume        = {84},
  number        = {17},
  pages         = {3760-3763},
  doi           = {10.1103/PhysRevLett.84.3760},
  archiveprefix = {arXiv},
  eprint        = {astro-ph/9909386},
  primaryclass  = {astro-ph},
  adsurl        = {https://ui.adsabs.harvard.edu/abs/2000PhRvL..84.3760S}
}

@article{2000PhRvL..85.4438A,
  author        = {{Armendariz-Picon}, C. and {Mukhanov}, V. and {Steinhardt}, Paul J.},
  title         = {{Dynamical Solution to the Problem of a Small Cosmological Constant and Late-Time Cosmic Acceleration}},
  journal       = {\prl},
  year          = 2000,
  month         = nov,
  volume        = {85},
  number        = {21},
  pages         = {4438-4441},
  doi           = {10.1103/PhysRevLett.85.4438},
  archiveprefix = {arXiv},
  eprint        = {astro-ph/0004134},
  primaryclass  = {astro-ph},
  adsurl        = {https://ui.adsabs.harvard.edu/abs/2000PhRvL..85.4438A}
}

@article{2001ApJ...547..574D,
  author        = {{Dav{\'e}}, Romeel and {Spergel}, David N. and {Steinhardt}, Paul J. and {Wandelt}, Benjamin D.},
  title         = {{Halo Properties in Cosmological Simulations of Self-interacting Cold Dark Matter}},
  journal       = {\apj},
  year          = 2001,
  month         = feb,
  volume        = {547},
  number        = {2},
  pages         = {574-589},
  doi           = {10.1086/318417},
  archiveprefix = {arXiv},
  eprint        = {astro-ph/0006218},
  primaryclass  = {astro-ph},
  adsurl        = {https://ui.adsabs.harvard.edu/abs/2001ApJ...547..574D}
}

@article{2001PhRvD..63j3510A,
  author        = {{Armendariz-Picon}, C. and {Mukhanov}, V. and {Steinhardt}, Paul J.},
  title         = {{Essentials of k-essence}},
  journal       = {\prd},
  year          = 2001,
  month         = may,
  volume        = {63},
  number        = {10},
  eid           = {103510},
  pages         = {103510},
  doi           = {10.1103/PhysRevD.63.103510},
  archiveprefix = {arXiv},
  eprint        = {astro-ph/0006373},
  primaryclass  = {astro-ph},
  adsurl        = {https://ui.adsabs.harvard.edu/abs/2001PhRvD..63j3510A}
}

@article{2001PhRvD..64f3501Z,
  author        = {{Zimdahl}, Winfried and {Schwarz}, Dominik J. and {Balakin}, Alexander B. and {Pav{\'o}n}, Diego},
  title         = {{Cosmic antifriction and accelerated expansion}},
  journal       = {\prd},
  year          = 2001,
  month         = sep,
  volume        = {64},
  number        = {6},
  eid           = {063501},
  pages         = {063501},
  doi           = {10.1103/PhysRevD.64.063501},
  archiveprefix = {arXiv},
  eprint        = {astro-ph/0009353},
  primaryclass  = {astro-ph},
  adsurl        = {https://ui.adsabs.harvard.edu/abs/2001PhRvD..64f3501Z}
}

@article{2003NJPh....5...85B,
  author        = {{Balakin}, Alexander B. and {Pav{\'o}n}, Diego and {Schwarz}, Dominik J. and {Zimdahl}, Winfried},
  title         = {{Curvature force and dark energy}},
  journal       = {New Journal of Physics},
  year          = 2003,
  month         = jul,
  volume        = {5},
  number        = {1},
  pages         = {85},
  doi           = {10.1088/1367-2630/5/1/385},
  archiveprefix = {arXiv},
  eprint        = {astro-ph/0302150},
  primaryclass  = {astro-ph},
  adsurl        = {https://ui.adsabs.harvard.edu/abs/2003NJPh....5...85B}
}

@article{2003PhRvL..91g1301C,
  author        = {{Caldwell}, Robert R. and {Kamionkowski}, Marc and {Weinberg}, Nevin N.},
  title         = {{Phantom Energy: Dark Energy with w<-1 Causes a Cosmic Doomsday}},
  journal       = {\prl},
  year          = 2003,
  month         = aug,
  volume        = {91},
  number        = {7},
  eid           = {071301},
  pages         = {071301},
  doi           = {10.1103/PhysRevLett.91.071301},
  archiveprefix = {arXiv},
  eprint        = {astro-ph/0302506},
  primaryclass  = {astro-ph},
  adsurl        = {https://ui.adsabs.harvard.edu/abs/2003PhRvL..91g1301C}
}

@article{2005PhRvD..72d3529L,
  author        = {{Linder}, Eric V.},
  title         = {{Cosmic growth history and expansion history}},
  journal       = {\prd},
  year          = 2005,
  month         = aug,
  volume        = {72},
  number        = {4},
  eid           = {043529},
  pages         = {043529},
  doi           = {10.1103/PhysRevD.72.043529},
  archiveprefix = {arXiv},
  eprint        = {astro-ph/0507263},
  primaryclass  = {astro-ph},
  adsurl        = {https://ui.adsabs.harvard.edu/abs/2005PhRvD..72d3529L}
}

@article{2008ApJ...679.1173R,
  author        = {{Randall}, Scott W. and {Markevitch}, Maxim and {Clowe}, Douglas and {Gonzalez}, Anthony H. and {Brada{\v{c}}}, Marusa},
  title         = {{Constraints on the Self-Interaction Cross Section of Dark Matter from Numerical Simulations of the Merging Galaxy Cluster 1E 0657-56}},
  journal       = {\apj},
  year          = 2008,
  month         = jun,
  volume        = {679},
  number        = {2},
  pages         = {1173-1180},
  doi           = {10.1086/587859},
  archiveprefix = {arXiv},
  eprint        = {0704.0261},
  primaryclass  = {astro-ph},
  adsurl        = {https://ui.adsabs.harvard.edu/abs/2008ApJ...679.1173R}
}

@article{2008CQGra..25t5006L,
  author        = {{Lima}, J.~A.~S. and {Silva}, F.~E. and {Santos}, R.~C.},
  title         = {{Accelerating cold dark matter cosmology ({\ensuremath{\Omega}}$_{{\ensuremath{\Lambda}}}$ {\ensuremath{\equiv}} 0)}},
  journal       = {Classical and Quantum Gravity},
  year          = 2008,
  month         = oct,
  volume        = {25},
  number        = {20},
  eid           = {205006},
  pages         = {205006},
  doi           = {10.1088/0264-9381/25/20/205006},
  archiveprefix = {arXiv},
  eprint        = {0807.3379},
  primaryclass  = {astro-ph},
  adsurl        = {https://ui.adsabs.harvard.edu/abs/2008CQGra..25t5006L}
}

@article{2008PhLB..669..266P,
  author        = {{Pereira}, S.~H. and {Lima}, J.~A.~S.},
  title         = {{On phantom thermodynamics}},
  journal       = {Physics Letters B},
  year          = 2008,
  month         = nov,
  volume        = {669},
  number        = {5},
  pages         = {266-270},
  doi           = {10.1016/j.physletb.2008.10.006},
  archiveprefix = {arXiv},
  eprint        = {0806.0682},
  primaryclass  = {gr-qc},
  adsurl        = {https://ui.adsabs.harvard.edu/abs/2008PhLB..669..266P}
}

@article{2009A&A...507...47B,
  author        = {{Basilakos}, S. and {Plionis}, M.},
  title         = {{Could dark matter interactions be an alternative to dark energy?}},
  journal       = {\aap},
  year          = 2009,
  month         = nov,
  volume        = {507},
  number        = {1},
  pages         = {47-52},
  doi           = {10.1051/0004-6361/200912661},
  archiveprefix = {arXiv},
  eprint        = {0807.4590},
  primaryclass  = {astro-ph},
  adsurl        = {https://ui.adsabs.harvard.edu/abs/2009A&A...507...47B}
}

@article{2010PhRvD..82b3504B,
  author        = {{Basilakos}, S. and {Lima}, J.~A.~S.},
  title         = {{Constraints on cold dark matter accelerating cosmologies and cluster formation}},
  journal       = {\prd},
  year          = 2010,
  month         = jul,
  volume        = {82},
  number        = {2},
  eid           = {023504},
  pages         = {023504},
  doi           = {10.1103/PhysRevD.82.023504},
  archiveprefix = {arXiv},
  eprint        = {1003.5754},
  primaryclass  = {astro-ph.CO},
  adsurl        = {https://ui.adsabs.harvard.edu/abs/2010PhRvD..82b3504B}
}

@article{2011A&A...529A..61C,
  author        = {{Cao}, S. and {Zhu}, Z.-H. and {Liang}, N.},
  title         = {{Observational constraints on interacting dark matter model without dark energy}},
  journal       = {\aap},
  year          = 2011,
  month         = may,
  volume        = {529},
  eid           = {A61},
  pages         = {A61},
  doi           = {10.1051/0004-6361/201016204},
  archiveprefix = {arXiv},
  eprint        = {1011.4848},
  primaryclass  = {astro-ph.CO},
  adsurl        = {https://ui.adsabs.harvard.edu/abs/2011A&A...529A..61C}
}

@article{2011PhRvD..84f3511J,
  author        = {{Jesus}, J.~F. and {Oliveira}, F.~A. and {Basilakos}, S. and {Lima}, J.~A.~S.},
  title         = {{Newtonian perturbations on models with matter creation}},
  journal       = {\prd},
  year          = 2011,
  month         = sep,
  volume        = {84},
  number        = {6},
  eid           = {063511},
  pages         = {063511},
  doi           = {10.1103/PhysRevD.84.063511},
  archiveprefix = {arXiv},
  eprint        = {1105.1027},
  primaryclass  = {astro-ph.CO},
  adsurl        = {https://ui.adsabs.harvard.edu/abs/2011PhRvD..84f3511J}
}

@article{2012MNRAS.422.1203B,
  author        = {{Boylan-Kolchin}, Michael and {Bullock}, James S. and {Kaplinghat}, Manoj},
  title         = {{The Milky Way's bright satellites as an apparent failure of {\ensuremath{\Lambda}}CDM}},
  journal       = {\mnras},
  year          = 2012,
  month         = may,
  volume        = {422},
  number        = {2},
  pages         = {1203-1218},
  doi           = {10.1111/j.1365-2966.2012.20695.x},
  archiveprefix = {arXiv},
  eprint        = {1111.2048},
  primaryclass  = {astro-ph.CO},
  adsurl        = {https://ui.adsabs.harvard.edu/abs/2012MNRAS.422.1203B}
}

@article{2012PhR...513....1C,
  author        = {{Clifton}, Timothy and {Ferreira}, Pedro G. and {Padilla}, Antonio and {Skordis}, Constantinos},
  title         = {{Modified gravity and cosmology}},
  journal       = {\physrep},
  year          = 2012,
  month         = mar,
  volume        = {513},
  number        = {1},
  pages         = {1-189},
  doi           = {10.1016/j.physrep.2012.01.001},
  archiveprefix = {arXiv},
  eprint        = {1106.2476},
  primaryclass  = {astro-ph.CO},
  adsurl        = {https://ui.adsabs.harvard.edu/abs/2012PhR...513....1C}
}

@article{2012PhRvD..86j3534L,
  author        = {{Lima}, J.~A.~S. and {Basilakos}, S. and {Costa}, F.~E.~M.},
  title         = {{New cosmic accelerating scenario without dark energy}},
  journal       = {\prd},
  year          = 2012,
  month         = nov,
  volume        = {86},
  number        = {10},
  eid           = {103534},
  pages         = {103534},
  doi           = {10.1103/PhysRevD.86.103534},
  archiveprefix = {arXiv},
  eprint        = {1205.0868},
  primaryclass  = {astro-ph.CO},
  adsurl        = {https://ui.adsabs.harvard.edu/abs/2012PhRvD..86j3534L}
}

@article{2013MNRAS.430..105P,
  author        = {{Peter}, Annika H.~G. and {Rocha}, Miguel and {Bullock}, James S. and {Kaplinghat}, Manoj},
  title         = {{Cosmological simulations with self-interacting dark matter - II. Halo shapes versus observations}},
  journal       = {\mnras},
  year          = 2013,
  month         = mar,
  volume        = {430},
  number        = {1},
  pages         = {105-120},
  doi           = {10.1093/mnras/sts535},
  archiveprefix = {arXiv},
  eprint        = {1208.3026},
  primaryclass  = {astro-ph.CO},
  adsurl        = {https://ui.adsabs.harvard.edu/abs/2013MNRAS.430..105P}
}

@article{2013PASP..125..306F,
  author        = {{Foreman-Mackey}, Daniel and {Hogg}, David W. and {Lang}, Dustin and {Goodman}, Jonathan},
  title         = {{emcee: The MCMC Hammer}},
  journal       = {\pasp},
  year          = 2013,
  month         = mar,
  volume        = {125},
  number        = {925},
  pages         = {306},
  doi           = {10.1086/670067},
  archiveprefix = {arXiv},
  eprint        = {1202.3665},
  primaryclass  = {astro-ph.IM},
  adsurl        = {https://ui.adsabs.harvard.edu/abs/2013PASP..125..306F}
}

@article{2013PhRvD..87k5007T,
  author        = {{Tulin}, Sean and {Yu}, Hai-Bo and {Zurek}, Kathryn M.},
  title         = {{Beyond collisionless dark matter: Particle physics dynamics for dark matter halo structure}},
  journal       = {\prd},
  year          = 2013,
  month         = jun,
  volume        = {87},
  number        = {11},
  eid           = {115007},
  pages         = {115007},
  doi           = {10.1103/PhysRevD.87.115007},
  archiveprefix = {arXiv},
  eprint        = {1302.3898},
  primaryclass  = {hep-ph},
  adsurl        = {https://ui.adsabs.harvard.edu/abs/2013PhRvD..87k5007T}
}

@article{2014JCAP...10..042L,
  author        = {{Lima}, J.~A.~S. and {Graef}, L.~L. and {Pav{\'o}n}, D. and {Basilakos}, Spyros},
  title         = {{Cosmic acceleration without dark energy: background tests and thermodynamic analysis}},
  journal       = {\jcap},
  year          = 2014,
  month         = oct,
  volume        = {2014},
  number        = {10},
  pages         = {042-042},
  doi           = {10.1088/1475-7516/2014/10/042},
  archiveprefix = {arXiv},
  eprint        = {1406.5538},
  primaryclass  = {gr-qc},
  adsurl        = {https://ui.adsabs.harvard.edu/abs/2014JCAP...10..042L}
}

@article{2015IJMPD..2450071C,
  author        = {{Cai}, Rong-Gen and {Guo}, Zong-Kuan and {Tang}, Bo},
  title         = {{Updated reduced CMB data and constraints on cosmological parameters}},
  journal       = {International Journal of Modern Physics D},
  year          = 2015,
  month         = jul,
  volume        = {24},
  number        = {10},
  eid           = {1550071},
  pages         = {1550071},
  doi           = {10.1142/S0218271815500716},
  archiveprefix = {arXiv},
  eprint        = {1409.0223},
  primaryclass  = {astro-ph.CO},
  adsurl        = {https://ui.adsabs.harvard.edu/abs/2015IJMPD..2450071C}
}

@article{2016MNRAS.459..673N,
  author        = {{Nunes}, Rafael C. and {Pan}, Supriya},
  title         = {{Cosmological consequences of an adiabatic matter creation process}},
  journal       = {\mnras},
  year          = 2016,
  month         = jun,
  volume        = {459},
  number        = {1},
  pages         = {673-682},
  doi           = {10.1093/mnras/stw566},
  archiveprefix = {arXiv},
  eprint        = {1603.02573},
  primaryclass  = {gr-qc},
  adsurl        = {https://ui.adsabs.harvard.edu/abs/2016MNRAS.459..673N}
}

@article{2016PhRvL.116d1302K,
  author        = {{Kaplinghat}, Manoj and {Tulin}, Sean and {Yu}, Hai-Bo},
  title         = {{Dark Matter Halos as Particle Colliders: Unified Solution to Small-Scale Structure Puzzles from Dwarfs to Clusters}},
  journal       = {\prl},
  year          = 2016,
  month         = jan,
  volume        = {116},
  number        = {4},
  eid           = {041302},
  pages         = {041302},
  doi           = {10.1103/PhysRevLett.116.041302},
  archiveprefix = {arXiv},
  eprint        = {1508.03339},
  primaryclass  = {astro-ph.CO},
  adsurl        = {https://ui.adsabs.harvard.edu/abs/2016PhRvL.116d1302K}
}

@article{2017A&A...606A..15C,
  author        = {{Cao}, Shuo and {Zheng}, Xiaogang and {Biesiada}, Marek and {Qi}, Jingzhao and {Chen}, Yun and {Zhu}, Zong-Hong},
  title         = {{Ultra-compact structure in intermediate-luminosity radio quasars: building a sample of standard cosmological rulers and improving the dark energy constraints up to z   3}},
  journal       = {\aap},
  year          = 2017,
  month         = sep,
  volume        = {606},
  eid           = {A15},
  pages         = {A15},
  doi           = {10.1051/0004-6361/201730551},
  archiveprefix = {arXiv},
  eprint        = {1708.08635},
  primaryclass  = {astro-ph.CO},
  adsurl        = {https://ui.adsabs.harvard.edu/abs/2017A&A...606A..15C}
}

@article{2017ARA&A..55..343B,
  author        = {{Bullock}, James S. and {Boylan-Kolchin}, Michael},
  title         = {{Small-Scale Challenges to the {\ensuremath{\Lambda}}CDM Paradigm}},
  journal       = {\araa},
  year          = 2017,
  month         = aug,
  volume        = {55},
  number        = {1},
  pages         = {343-387},
  doi           = {10.1146/annurev-astro-091916-055313},
  archiveprefix = {arXiv},
  eprint        = {1707.04256},
  primaryclass  = {astro-ph.CO},
  adsurl        = {https://ui.adsabs.harvard.edu/abs/2017ARA&A..55..343B}
}

@article{2017arXiv170403472H,
  author        = {{Heavens}, Alan and {Fantaye}, Yabebal and {Mootoovaloo}, Arrykrishna and {Eggers}, Hans and {Hosenie}, Zafiirah and {Kroon}, Steve and {Sellentin}, Elena},
  title         = {{Marginal Likelihoods from Monte Carlo Markov Chains}},
  journal       = {arXiv e-prints},
  year          = 2017,
  month         = apr,
  eid           = {arXiv:1704.03472},
  pages         = {arXiv:1704.03472},
  doi           = {10.48550/arXiv.1704.03472},
  archiveprefix = {arXiv},
  eprint        = {1704.03472},
  primaryclass  = {stat.CO},
  adsurl        = {https://ui.adsabs.harvard.edu/abs/2017arXiv170403472H}
}

@article{2017PhRvD..96b3542N,
  author        = {{Nesseris}, Savvas and {Pantazis}, George and {Perivolaropoulos}, Leandros},
  title         = {{Tension and constraints on modified gravity parametrizations of G$_{eff}$(z ) from growth rate and Planck data}},
  journal       = {\prd},
  year          = 2017,
  month         = jul,
  volume        = {96},
  number        = {2},
  eid           = {023542},
  pages         = {023542},
  doi           = {10.1103/PhysRevD.96.023542},
  archiveprefix = {arXiv},
  eprint        = {1703.10538},
  primaryclass  = {astro-ph.CO},
  adsurl        = {https://ui.adsabs.harvard.edu/abs/2017PhRvD..96b3542N}
}

@article{2018ApJ...853..109E,
  author        = {{Elbert}, Oliver D. and {Bullock}, James S. and {Kaplinghat}, Manoj and {Garrison-Kimmel}, Shea and {Graus}, Andrew S. and {Rocha}, Miguel},
  title         = {{A Testable Conspiracy: Simulating Baryonic Effects on Self-interacting Dark Matter Halos}},
  journal       = {\apj},
  year          = 2018,
  month         = feb,
  volume        = {853},
  number        = {2},
  eid           = {109},
  pages         = {109},
  doi           = {10.3847/1538-4357/aa9710},
  archiveprefix = {arXiv},
  eprint        = {1609.08626},
  primaryclass  = {astro-ph.GA},
  adsurl        = {https://ui.adsabs.harvard.edu/abs/2018ApJ...853..109E}
}

@article{2018ApJ...869..104W,
  author        = {{Wittman}, David and {Golovich}, Nathan and {Dawson}, William A.},
  title         = {{The Mismeasure of Mergers: Revised Limits on Self-interacting Dark Matter in Merging Galaxy Clusters}},
  journal       = {\apj},
  year          = 2018,
  month         = dec,
  volume        = {869},
  number        = {2},
  eid           = {104},
  pages         = {104},
  doi           = {10.3847/1538-4357/aaee77},
  archiveprefix = {arXiv},
  eprint        = {1701.05877},
  primaryclass  = {astro-ph.CO},
  adsurl        = {https://ui.adsabs.harvard.edu/abs/2018ApJ...869..104W}
}

@article{2018PhR...730....1T,
  author        = {{Tulin}, Sean and {Yu}, Hai-Bo},
  title         = {{Dark matter self-interactions and small scale structure}},
  journal       = {\physrep},
  year          = 2018,
  month         = feb,
  volume        = {730},
  pages         = {1-57},
  doi           = {10.1016/j.physrep.2017.11.004},
  archiveprefix = {arXiv},
  eprint        = {1705.02358},
  primaryclass  = {hep-ph},
  adsurl        = {https://ui.adsabs.harvard.edu/abs/2018PhR...730....1T}
}

@article{2019ApJ...873..111I,
  author        = {{Ivezi{\'c}}, {\v{Z}}eljko and {Kahn}, Steven M. and {Tyson}, J. Anthony and {Abel}, Bob and {Acosta}, Emily and {Allsman}, Robyn and {Alonso}, David and {AlSayyad}, Yusra and {Anderson}, Scott F. and {Andrew}, John and {Angel}, James Roger P. and {Angeli}, George Z. and {Ansari}, Reza and {Antilogus}, Pierre and {Araujo}, Constanza and {Armstrong}, Robert and {Arndt}, Kirk T. and {Astier}, Pierre and {Aubourg}, {\'E}ric and {Auza}, Nicole and {Axelrod}, Tim S. and {Bard}, Deborah J. and {Barr}, Jeff D. and {Barrau}, Aurelian and {Bartlett}, James G. and {Bauer}, Amanda E. and {Bauman}, Brian J. and {Baumont}, Sylvain and {Bechtol}, Ellen and {Bechtol}, Keith and {Becker}, Andrew C. and {Becla}, Jacek and {Beldica}, Cristina and {Bellavia}, Steve and {Bianco}, Federica B. and {Biswas}, Rahul and {Blanc}, Guillaume and {Blazek}, Jonathan and {Blandford}, Roger D. and {Bloom}, Josh S. and {Bogart}, Joanne and {Bond}, Tim W. and {Booth}, Michael T. and {Borgland}, Anders W. and {Borne}, Kirk and {Bosch}, James F. and {Boutigny}, Dominique and {Brackett}, Craig A. and {Bradshaw}, Andrew and {Brandt}, William Nielsen and {Brown}, Michael E. and {Bullock}, James S. and {Burchat}, Patricia and {Burke}, David L. and {Cagnoli}, Gianpietro and {Calabrese}, Daniel and {Callahan}, Shawn and {Callen}, Alice L. and {Carlin}, Jeffrey L. and {Carlson}, Erin L. and {Chandrasekharan}, Srinivasan and {Charles-Emerson}, Glenaver and {Chesley}, Steve and {Cheu}, Elliott C. and {Chiang}, Hsin-Fang and {Chiang}, James and {Chirino}, Carol and {Chow}, Derek and {Ciardi}, David R. and {Claver}, Charles F. and {Cohen-Tanugi}, Johann and {Cockrum}, Joseph J. and {Coles}, Rebecca and {Connolly}, Andrew J. and {Cook}, Kem H. and {Cooray}, Asantha and {Covey}, Kevin R. and {Cribbs}, Chris and {Cui}, Wei and {Cutri}, Roc and {Daly}, Philip N. and {Daniel}, Scott F. and {Daruich}, Felipe and {Daubard}, Guillaume and {Daues}, Greg and {Dawson}, William and {Delgado}, Francisco and {Dellapenna}, Alfred and {de Peyster}, Robert and {de Val-Borro}, Miguel and {Digel}, Seth W. and {Doherty}, Peter and {Dubois}, Richard and {Dubois-Felsmann}, Gregory P. and {Durech}, Josef and {Economou}, Frossie and {Eifler}, Tim and {Eracleous}, Michael and {Emmons}, Benjamin L. and {Fausti Neto}, Angelo and {Ferguson}, Henry and {Figueroa}, Enrique and {Fisher-Levine}, Merlin and {Focke}, Warren and {Foss}, Michael D. and {Frank}, James and {Freemon}, Michael D. and {Gangler}, Emmanuel and {Gawiser}, Eric and {Geary}, John C. and {Gee}, Perry and {Geha}, Marla and {Gessner}, Charles J.~B. and {Gibson}, Robert R. and {Gilmore}, D. Kirk and {Glanzman}, Thomas and {Glick}, William and {Goldina}, Tatiana and {Goldstein}, Daniel A. and {Goodenow}, Iain and {Graham}, Melissa L. and {Gressler}, William J. and {Gris}, Philippe and {Guy}, Leanne P. and {Guyonnet}, Augustin and {Haller}, Gunther and {Harris}, Ron and {Hascall}, Patrick A. and {Haupt}, Justine and {Hernandez}, Fabio and {Herrmann}, Sven and {Hileman}, Edward and {Hoblitt}, Joshua and {Hodgson}, John A. and {Hogan}, Craig and {Howard}, James D. and {Huang}, Dajun and {Huffer}, Michael E. and {Ingraham}, Patrick and {Innes}, Walter R. and {Jacoby}, Suzanne H. and {Jain}, Bhuvnesh and {Jammes}, Fabrice and {Jee}, M. James and {Jenness}, Tim and {Jernigan}, Garrett and {Jevremovi{\'c}}, Darko and {Johns}, Kenneth and {Johnson}, Anthony S. and {Johnson}, Margaret W.~G. and {Jones}, R. Lynne and {Juramy-Gilles}, Claire and {Juri{\'c}}, Mario and {Kalirai}, Jason S. and {Kallivayalil}, Nitya J. and {Kalmbach}, Bryce and {Kantor}, Jeffrey P. and {Karst}, Pierre and {Kasliwal}, Mansi M. and {Kelly}, Heather and {Kessler}, Richard and {Kinnison}, Veronica and {Kirkby}, David and {Knox}, Lloyd and {Kotov}, Ivan V. and {Krabbendam}, Victor L. and {Krughoff}, K. Simon and {Kub{\'a}nek}, Petr and {Kuczewski}, John and {Kulkarni}, Shri and {Ku}, John and {Kurita}, Nadine R. and {Lage}, Craig S. and {Lambert}, Ron and {Lange}, Travis and {Langton}, J. Brian and {Le Guillou}, Laurent and {Levine}, Deborah and {Liang}, Ming and {Lim}, Kian-Tat and {Lintott}, Chris J. and {Long}, Kevin E. and {Lopez}, Margaux and {Lotz}, Paul J. and {Lupton}, Robert H. and {Lust}, Nate B. and {MacArthur}, Lauren A. and {Mahabal}, Ashish and {Mandelbaum}, Rachel and {Markiewicz}, Thomas W. and {Marsh}, Darren S. and {Marshall}, Philip J. and {Marshall}, Stuart and {May}, Morgan and {McKercher}, Robert and {McQueen}, Michelle and {Meyers}, Joshua and {Migliore}, Myriam and {Miller}, Michelle and {Mills}, David J.},
  title         = {{LSST: From Science Drivers to Reference Design and Anticipated Data Products}},
  journal       = {\apj},
  year          = 2019,
  month         = mar,
  volume        = {873},
  number        = {2},
  eid           = {111},
  pages         = {111},
  doi           = {10.3847/1538-4357/ab042c},
  archiveprefix = {arXiv},
  eprint        = {0805.2366},
  primaryclass  = {astro-ph},
  adsurl        = {https://ui.adsabs.harvard.edu/abs/2019ApJ...873..111I}
}

@article{2019ApJ...887..163M,
  author        = {{Ma}, Yu-Bo and {Cao}, Shuo and {Zhang}, Jia and {Qi}, Jingzhao and {Liu}, Tonghua and {Liu}, Yuting and {Geng}, Shuaibo},
  title         = {{Testing Cosmic Opacity with the Combination of Strongly Lensed and Unlensed Supernova Ia}},
  journal       = {\apj},
  year          = 2019,
  month         = dec,
  volume        = {887},
  number        = {2},
  eid           = {163},
  pages         = {163},
  doi           = {10.3847/1538-4357/ab50c4},
  archiveprefix = {arXiv},
  eprint        = {1910.10351},
  primaryclass  = {astro-ph.CO},
  adsurl        = {https://ui.adsabs.harvard.edu/abs/2019ApJ...887..163M}
}

@article{2019JCAP...02..028C,
  author        = {{Chen}, Lu and {Huang}, Qing-Guo and {Wang}, Ke},
  title         = {{Distance priors from Planck final release}},
  journal       = {\jcap},
  year          = 2019,
  month         = feb,
  volume        = {2019},
  number        = {2},
  eid           = {028},
  pages         = {028},
  doi           = {10.1088/1475-7516/2019/02/028},
  archiveprefix = {arXiv},
  eprint        = {1808.05724},
  primaryclass  = {astro-ph.CO},
  adsurl        = {https://ui.adsabs.harvard.edu/abs/2019JCAP...02..028C}
}

@article{2019JCAP...02..056A,
  author        = {{Ade}, Peter and {Aguirre}, James and {Ahmed}, Zeeshan and {Aiola}, Simone and {Ali}, Aamir and {Alonso}, David and {Alvarez}, Marcelo A. and {Arnold}, Kam and {Ashton}, Peter and {Austermann}, Jason and {Awan}, Humna and {Baccigalupi}, Carlo and {Baildon}, Taylor and {Barron}, Darcy and {Battaglia}, Nick and {Battye}, Richard and {Baxter}, Eric and {Bazarko}, Andrew and {Beall}, James A. and {Bean}, Rachel and {Beck}, Dominic and {Beckman}, Shawn and {Beringue}, Benjamin and {Bianchini}, Federico and {Boada}, Steven and {Boettger}, David and {Bond}, J. Richard and {Borrill}, Julian and {Brown}, Michael L. and {Bruno}, Sarah Marie and {Bryan}, Sean and {Calabrese}, Erminia and {Calafut}, Victoria and {Calisse}, Paolo and {Carron}, Julien and {Challinor}, Anthony and {Chesmore}, Grace and {Chinone}, Yuji and {Chluba}, Jens and {Cho}, Hsiao-Mei Sherry and {Choi}, Steve and {Coppi}, Gabriele and {Cothard}, Nicholas F. and {Coughlin}, Kevin and {Crichton}, Devin and {Crowley}, Kevin D. and {Crowley}, Kevin T. and {Cukierman}, Ari and {D'Ewart}, John M. and {D{\"u}nner}, Rolando and {de Haan}, Tijmen and {Devlin}, Mark and {Dicker}, Simon and {Didier}, Joy and {Dobbs}, Matt and {Dober}, Bradley and {Duell}, Cody J. and {Duff}, Shannon and {Duivenvoorden}, Adri and {Dunkley}, Jo and {Dusatko}, John and {Errard}, Josquin and {Fabbian}, Giulio and {Feeney}, Stephen and {Ferraro}, Simone and {Flux{\`a}}, Pedro and {Freese}, Katherine and {Frisch}, Josef C. and {Frolov}, Andrei and {Fuller}, George and {Fuzia}, Brittany and {Galitzki}, Nicholas and {Gallardo}, Patricio A. and {Tomas Galvez Ghersi}, Jose and {Gao}, Jiansong and {Gawiser}, Eric and {Gerbino}, Martina and {Gluscevic}, Vera and {Goeckner-Wald}, Neil and {Golec}, Joseph and {Gordon}, Sam and {Gralla}, Megan and {Green}, Daniel and {Grigorian}, Arpi and {Groh}, John and {Groppi}, Chris and {Guan}, Yilun and {Gudmundsson}, Jon E. and {Han}, Dongwon and {Hargrave}, Peter and {Hasegawa}, Masaya and {Hasselfield}, Matthew and {Hattori}, Makoto and {Haynes}, Victor and {Hazumi}, Masashi and {He}, Yizhou and {Healy}, Erin and {Henderson}, Shawn W. and {Hervias-Caimapo}, Carlos and {Hill}, Charles A. and {Hill}, J. Colin and {Hilton}, Gene and {Hilton}, Matt and {Hincks}, Adam D. and {Hinshaw}, Gary and {Hlo{\v{z}}ek}, Ren{\'e}e and {Ho}, Shirley and {Ho}, Shuay-Pwu Patty and {Howe}, Logan and {Huang}, Zhiqi and {Hubmayr}, Johannes and {Huffenberger}, Kevin and {Hughes}, John P. and {Ijjas}, Anna and {Ikape}, Margaret and {Irwin}, Kent and {Jaffe}, Andrew H. and {Jain}, Bhuvnesh and {Jeong}, Oliver and {Kaneko}, Daisuke and {Karpel}, Ethan D. and {Katayama}, Nobuhiko and {Keating}, Brian and {Kernasovskiy}, Sarah S. and {Keskitalo}, Reijo and {Kisner}, Theodore and {Kiuchi}, Kenji and {Klein}, Jeff and {Knowles}, Kenda and {Koopman}, Brian and {Kosowsky}, Arthur and {Krachmalnicoff}, Nicoletta and {Kuenstner}, Stephen E. and {Kuo}, Chao-Lin and {Kusaka}, Akito and {Lashner}, Jacob and {Lee}, Adrian and {Lee}, Eunseong and {Leon}, David and {Leung}, Jason S.-Y. and {Lewis}, Antony and {Li}, Yaqiong and {Li}, Zack and {Limon}, Michele and {Linder}, Eric and {Lopez-Caraballo}, Carlos and {Louis}, Thibaut and {Lowry}, Lindsay and {Lungu}, Marius and {Madhavacheril}, Mathew and {Mak}, Daisy and {Maldonado}, Felipe and {Mani}, Hamdi and {Mates}, Ben and {Matsuda}, Frederick and {Maurin}, Lo{\"\i}c and {Mauskopf}, Phil and {May}, Andrew and {McCallum}, Nialh and {McKenney}, Chris and {McMahon}, Jeff and {Meerburg}, P. Daniel and {Meyers}, Joel and {Miller}, Amber and {Mirmelstein}, Mark and {Moodley}, Kavilan and {Munchmeyer}, Moritz and {Munson}, Charles and {Naess}, Sigurd and {Nati}, Federico and {Navaroli}, Martin and {Newburgh}, Laura and {Nguyen}, Ho Nam and {Niemack}, Michael and {Nishino}, Haruki and {Orlowski-Scherer}, John and {Page}, Lyman and {Partridge}, Bruce and {Peloton}, Julien and {Perrotta}, Francesca and {Piccirillo}, Lucio and {Pisano}, Giampaolo and {Poletti}, Davide and {Puddu}, Roberto and {Puglisi}, Giuseppe and {Raum}, Chris and {Reichardt}, Christian L. and {Remazeilles}, Mathieu and {Rephaeli}, Yoel and {Riechers}, Dominik and {Rojas}, Felipe and {Roy}, Anirban and {Sadeh}, Sharon and {Sakurai}, Yuki and {Salatino}, Maria and {Sathyanarayana Rao}, Mayuri and {Schaan}, Emmanuel and {Schmittfull}, Marcel and {Sehgal}, Neelima and {Seibert}, Joseph},
  title         = {{The Simons Observatory: science goals and forecasts}},
  journal       = {\jcap},
  year          = 2019,
  month         = feb,
  volume        = {2019},
  number        = {2},
  eid           = {056},
  pages         = {056},
  doi           = {10.1088/1475-7516/2019/02/056},
  archiveprefix = {arXiv},
  eprint        = {1808.07445},
  primaryclass  = {astro-ph.CO},
  adsurl        = {https://ui.adsabs.harvard.edu/abs/2019JCAP...02..056A}
}

@article{2019PhRvL.122g1103C,
  author        = {{Chu}, Xiaoyong and {Garcia-Cely}, Camilo and {Murayama}, Hitoshi},
  title         = {{Velocity Dependence from Resonant Self-Interacting Dark Matter}},
  journal       = {\prl},
  year          = 2019,
  month         = feb,
  volume        = {122},
  number        = {7},
  eid           = {071103},
  pages         = {071103},
  doi           = {10.1103/PhysRevLett.122.071103},
  archiveprefix = {arXiv},
  eprint        = {1810.04709},
  primaryclass  = {hep-ph},
  adsurl        = {https://ui.adsabs.harvard.edu/abs/2019PhRvL.122g1103C}
}

@article{2020A&A...641A...6P,
  author        = {{Planck Collaboration} and {Aghanim}, N. and {Akrami}, Y. and {Ashdown}, M. and {Aumont}, J. and {Baccigalupi}, C. and {Ballardini}, M. and {Banday}, A.~J. and {Barreiro}, R.~B. and {Bartolo}, N. and {Basak}, S. and {Battye}, R. and {Benabed}, K. and {Bernard}, J.-P. and {Bersanelli}, M. and {Bielewicz}, P. and {Bock}, J.~J. and {Bond}, J.~R. and {Borrill}, J. and {Bouchet}, F.~R. and {Boulanger}, F. and {Bucher}, M. and {Burigana}, C. and {Butler}, R.~C. and {Calabrese}, E. and {Cardoso}, J.-F. and {Carron}, J. and {Challinor}, A. and {Chiang}, H.~C. and {Chluba}, J. and {Colombo}, L.~P.~L. and {Combet}, C. and {Contreras}, D. and {Crill}, B.~P. and {Cuttaia}, F. and {de Bernardis}, P. and {de Zotti}, G. and {Delabrouille}, J. and {Delouis}, J.-M. and {Di Valentino}, E. and {Diego}, J.~M. and {Dor{\'e}}, O. and {Douspis}, M. and {Ducout}, A. and {Dupac}, X. and {Dusini}, S. and {Efstathiou}, G. and {Elsner}, F. and {En{\ss}lin}, T.~A. and {Eriksen}, H.~K. and {Fantaye}, Y. and {Farhang}, M. and {Fergusson}, J. and {Fernandez-Cobos}, R. and {Finelli}, F. and {Forastieri}, F. and {Frailis}, M. and {Fraisse}, A.~A. and {Franceschi}, E. and {Frolov}, A. and {Galeotta}, S. and {Galli}, S. and {Ganga}, K. and {G{\'e}nova-Santos}, R.~T. and {Gerbino}, M. and {Ghosh}, T. and {Gonz{\'a}lez-Nuevo}, J. and {G{\'o}rski}, K.~M. and {Gratton}, S. and {Gruppuso}, A. and {Gudmundsson}, J.~E. and {Hamann}, J. and {Handley}, W. and {Hansen}, F.~K. and {Herranz}, D. and {Hildebrandt}, S.~R. and {Hivon}, E. and {Huang}, Z. and {Jaffe}, A.~H. and {Jones}, W.~C. and {Karakci}, A. and {Keih{\"a}nen}, E. and {Keskitalo}, R. and {Kiiveri}, K. and {Kim}, J. and {Kisner}, T.~S. and {Knox}, L. and {Krachmalnicoff}, N. and {Kunz}, M. and {Kurki-Suonio}, H. and {Lagache}, G. and {Lamarre}, J.-M. and {Lasenby}, A. and {Lattanzi}, M. and {Lawrence}, C.~R. and {Le Jeune}, M. and {Lemos}, P. and {Lesgourgues}, J. and {Levrier}, F. and {Lewis}, A. and {Liguori}, M. and {Lilje}, P.~B. and {Lilley}, M. and {Lindholm}, V. and {L{\'o}pez-Caniego}, M. and {Lubin}, P.~M. and {Ma}, Y.-Z. and {Mac{\'\i}as-P{\'e}rez}, J.~F. and {Maggio}, G. and {Maino}, D. and {Mandolesi}, N. and {Mangilli}, A. and {Marcos-Caballero}, A. and {Maris}, M. and {Martin}, P.~G. and {Martinelli}, M. and {Mart{\'\i}nez-Gonz{\'a}lez}, E. and {Matarrese}, S. and {Mauri}, N. and {McEwen}, J.~D. and {Meinhold}, P.~R. and {Melchiorri}, A. and {Mennella}, A. and {Migliaccio}, M. and {Millea}, M. and {Mitra}, S. and {Miville-Desch{\^e}nes}, M.-A. and {Molinari}, D. and {Montier}, L. and {Morgante}, G. and {Moss}, A. and {Natoli}, P. and {N{\o}rgaard-Nielsen}, H.~U. and {Pagano}, L. and {Paoletti}, D. and {Partridge}, B. and {Patanchon}, G. and {Peiris}, H.~V. and {Perrotta}, F. and {Pettorino}, V. and {Piacentini}, F. and {Polastri}, L. and {Polenta}, G. and {Puget}, J.-L. and {Rachen}, J.~P. and {Reinecke}, M. and {Remazeilles}, M. and {Renzi}, A. and {Rocha}, G. and {Rosset}, C. and {Roudier}, G. and {Rubi{\~n}o-Mart{\'\i}n}, J.~A. and {Ruiz-Granados}, B. and {Salvati}, L. and {Sandri}, M. and {Savelainen}, M. and {Scott}, D. and {Shellard}, E.~P.~S. and {Sirignano}, C. and {Sirri}, G. and {Spencer}, L.~D. and {Sunyaev}, R. and {Suur-Uski}, A.-S. and {Tauber}, J.~A. and {Tavagnacco}, D. and {Tenti}, M. and {Toffolatti}, L. and {Tomasi}, M. and {Trombetti}, T. and {Valenziano}, L. and {Valiviita}, J. and {Van Tent}, B. and {Vibert}, L. and {Vielva}, P. and {Villa}, F. and {Vittorio}, N. and {Wandelt}, B.~D. and {Wehus}, I.~K. and {White}, M. and {White}, S.~D.~M. and {Zacchei}, A. and {Zonca}, A.},
  title         = {{Planck 2018 results. VI. Cosmological parameters}},
  journal       = {\aap},
  year          = 2020,
  month         = sep,
  volume        = {641},
  eid           = {A6},
  pages         = {A6},
  doi           = {10.1051/0004-6361/201833910},
  archiveprefix = {arXiv},
  eprint        = {1807.06209},
  primaryclass  = {astro-ph.CO},
  adsurl        = {https://ui.adsabs.harvard.edu/abs/2020A&A...641A...6P}
}

@article{2020ApJ...901..129L,
  author        = {{Liu}, Yuting and {Cao}, Shuo and {Liu}, Tonghua and {Li}, Xiaolei and {Geng}, Shuaibo and {Lian}, Yujie and {Guo}, Wuzheng},
  title         = {{Model-independent Constraints on Cosmic Curvature: Implication from Updated Hubble Diagram of High-redshift Standard Candles}},
  journal       = {\apj},
  year          = 2020,
  month         = oct,
  volume        = {901},
  number        = {2},
  eid           = {129},
  pages         = {129},
  doi           = {10.3847/1538-4357/abb0e4},
  archiveprefix = {arXiv},
  eprint        = {2008.08378},
  primaryclass  = {astro-ph.CO},
  adsurl        = {https://ui.adsabs.harvard.edu/abs/2020ApJ...901..129L}
}

@article{2020PhRvL.125k1105Y,
  author        = {{Yang}, Daneng and {Yu}, Hai-Bo and {An}, Haipeng},
  title         = {{Self-Interacting Dark Matter and the Origin of Ultradiffuse Galaxies NGC1052-DF2 and -DF4}},
  journal       = {\prl},
  year          = 2020,
  month         = sep,
  volume        = {125},
  number        = {11},
  eid           = {111105},
  pages         = {111105},
  doi           = {10.1103/PhysRevLett.125.111105},
  archiveprefix = {arXiv},
  eprint        = {2002.02102},
  primaryclass  = {astro-ph.GA},
  adsurl        = {https://ui.adsabs.harvard.edu/abs/2020PhRvL.125k1105Y}
}

@article{2021EPJC...81..903L,
  author        = {{Liu}, Tonghua and {Cao}, Shuo and {Zhang}, Sixuan and {Gong}, Xiaolong and {Guo}, Wuzheng and {Zheng}, Chenfa},
  title         = {{Revisiting the cosmic distance duality relation with machine learning reconstruction methods: the combination of HII galaxies and ultra-compact radio quasars}},
  journal       = {European Physical Journal C},
  year          = 2021,
  month         = oct,
  volume        = {81},
  number        = {10},
  eid           = {903},
  pages         = {903},
  doi           = {10.1140/epjc/s10052-021-09713-5},
  archiveprefix = {arXiv},
  eprint        = {2110.00927},
  primaryclass  = {astro-ph.CO},
  adsurl        = {https://ui.adsabs.harvard.edu/abs/2021EPJC...81..903L}
}

@article{2021MNRAS.502L..16C,
  author        = {{Cao}, Shuo and {Qi}, Jingzhao and {Biesiada}, Marek and {Liu}, Tonghua and {Li}, Jin and {Zhu}, Zong-Hong},
  title         = {{Measuring the viscosity of dark matter with strongly lensed gravitational waves}},
  journal       = {\mnras},
  year          = 2021,
  month         = mar,
  volume        = {502},
  number        = {1},
  pages         = {L16-L20},
  doi           = {10.1093/mnrasl/slaa205},
  archiveprefix = {arXiv},
  eprint        = {2012.12462},
  primaryclass  = {astro-ph.CO},
  adsurl        = {https://ui.adsabs.harvard.edu/abs/2021MNRAS.502L..16C}
}

@article{2021MNRAS.506.2181L,
  author        = {{Liu}, Tonghua and {Cao}, Shuo and {Biesiada}, Marek and {Liu}, Yuting and {Lian}, Yujie and {Zhang}, Yilong},
  title         = {{Consistency testing for invariance of the speed of light at different redshifts: the newest results from strong lensing and Type Ia supernovae observations}},
  journal       = {\mnras},
  year          = 2021,
  month         = sep,
  volume        = {506},
  number        = {2},
  pages         = {2181-2188},
  doi           = {10.1093/mnras/stab1868},
  archiveprefix = {arXiv},
  eprint        = {2106.15145},
  primaryclass  = {astro-ph.CO},
  adsurl        = {https://ui.adsabs.harvard.edu/abs/2021MNRAS.506.2181L}
}

@article{2021PhRvD.103b3017H,
  author        = {{Hayashi}, Kohei and {Ibe}, Masahiro and {Kobayashi}, Shin and {Nakayama}, Yuhei and {Shirai}, Satoshi},
  title         = {{Probing dark matter self-interaction with ultrafaint dwarf galaxies}},
  journal       = {\prd},
  year          = 2021,
  month         = jan,
  volume        = {103},
  number        = {2},
  eid           = {023017},
  pages         = {023017},
  doi           = {10.1103/PhysRevD.103.023017},
  archiveprefix = {arXiv},
  eprint        = {2008.02529},
  primaryclass  = {astro-ph.CO},
  adsurl        = {https://ui.adsabs.harvard.edu/abs/2021PhRvD.103b3017H}
}

@article{2022A&A...659L...5C,
  author        = {{Cao}, Shuo and {Qi}, Jingzhao and {Cao}, Zhoujian and {Biesiada}, Marek and {Cheng}, Wei and {Zhu}, Zong-Hong},
  title         = {{Direct measurement of the distribution of dark matter with strongly lensed gravitational waves}},
  journal       = {\aap},
  year          = 2022,
  month         = mar,
  volume        = {659},
  eid           = {L5},
  pages         = {L5},
  doi           = {10.1051/0004-6361/202142694},
  archiveprefix = {arXiv},
  eprint        = {2202.08714},
  primaryclass  = {astro-ph.CO},
  adsurl        = {https://ui.adsabs.harvard.edu/abs/2022A&A...659L...5C}
}

@article{2022ApJ...934L...7R,
  author        = {{Riess}, Adam G. and {Yuan}, Wenlong and {Macri}, Lucas M. and {Scolnic}, Dan and {Brout}, Dillon and {Casertano}, Stefano and {Jones}, David O. and {Murakami}, Yukei and {Anand}, Gagandeep S. and {Breuval}, Louise and {Brink}, Thomas G. and {Filippenko}, Alexei V. and {Hoffmann}, Samantha and {Jha}, Saurabh W. and {D'arcy Kenworthy}, W. and {Mackenty}, John and {Stahl}, Benjamin E. and {Zheng}, WeiKang},
  title         = {{A Comprehensive Measurement of the Local Value of the Hubble Constant with 1 km s$^{-1}$ Mpc$^{-1}$ Uncertainty from the Hubble Space Telescope and the SH0ES Team}},
  journal       = {\apjl},
  year          = 2022,
  month         = jul,
  volume        = {934},
  number        = {1},
  eid           = {L7},
  pages         = {L7},
  doi           = {10.3847/2041-8213/ac5c5b},
  archiveprefix = {arXiv},
  eprint        = {2112.04510},
  primaryclass  = {astro-ph.CO},
  adsurl        = {https://ui.adsabs.harvard.edu/abs/2022ApJ...934L...7R}
}

@article{2022ApJ...938..110B,
  author        = {{Brout}, Dillon and {Scolnic}, Dan and {Popovic}, Brodie and {Riess}, Adam G. and {Carr}, Anthony and {Zuntz}, Joe and {Kessler}, Rick and {Davis}, Tamara M. and {Hinton}, Samuel and {Jones}, David and {Kenworthy}, W. D'Arcy and {Peterson}, Erik R. and {Said}, Khaled and {Taylor}, Georgie and {Ali}, Noor and {Armstrong}, Patrick and {Charvu}, Pranav and {Dwomoh}, Arianna and {Meldorf}, Cole and {Palmese}, Antonella and {Qu}, Helen and {Rose}, Benjamin M. and {Sanchez}, Bruno and {Stubbs}, Christopher W. and {Vincenzi}, Maria and {Wood}, Charlotte M. and {Brown}, Peter J. and {Chen}, Rebecca and {Chambers}, Ken and {Coulter}, David A. and {Dai}, Mi and {Dimitriadis}, Georgios and {Filippenko}, Alexei V. and {Foley}, Ryan J. and {Jha}, Saurabh W. and {Kelsey}, Lisa and {Kirshner}, Robert P. and {M{\"o}ller}, Anais and {Muir}, Jessie and {Nadathur}, Seshadri and {Pan}, Yen-Chen and {Rest}, Armin and {Rojas-Bravo}, Cesar and {Sako}, Masao and {Siebert}, Matthew R. and {Smith}, Mat and {Stahl}, Benjamin E. and {Wiseman}, Phil},
  title         = {{The Pantheon+ Analysis: Cosmological Constraints}},
  journal       = {\apj},
  year          = 2022,
  month         = oct,
  volume        = {938},
  number        = {2},
  eid           = {110},
  pages         = {110},
  doi           = {10.3847/1538-4357/ac8e04},
  archiveprefix = {arXiv},
  eprint        = {2202.04077},
  primaryclass  = {astro-ph.CO},
  adsurl        = {https://ui.adsabs.harvard.edu/abs/2022ApJ...938..110B}
}

@article{2022ApJ...938..113S,
  author        = {{Scolnic}, Dan and {Brout}, Dillon and {Carr}, Anthony and {Riess}, Adam G. and {Davis}, Tamara M. and {Dwomoh}, Arianna and {Jones}, David O. and {Ali}, Noor and {Charvu}, Pranav and {Chen}, Rebecca and {Peterson}, Erik R. and {Popovic}, Brodie and {Rose}, Benjamin M. and {Wood}, Charlotte M. and {Brown}, Peter J. and {Chambers}, Ken and {Coulter}, David A. and {Dettman}, Kyle G. and {Dimitriadis}, Georgios and {Filippenko}, Alexei V. and {Foley}, Ryan J. and {Jha}, Saurabh W. and {Kilpatrick}, Charles D. and {Kirshner}, Robert P. and {Pan}, Yen-Chen and {Rest}, Armin and {Rojas-Bravo}, Cesar and {Siebert}, Matthew R. and {Stahl}, Benjamin E. and {Zheng}, WeiKang},
  title         = {{The Pantheon+ Analysis: The Full Data Set and Light-curve Release}},
  journal       = {\apj},
  year          = 2022,
  month         = oct,
  volume        = {938},
  number        = {2},
  eid           = {113},
  pages         = {113},
  doi           = {10.3847/1538-4357/ac8b7a},
  archiveprefix = {arXiv},
  eprint        = {2112.03863},
  primaryclass  = {astro-ph.CO},
  adsurl        = {https://ui.adsabs.harvard.edu/abs/2022ApJ...938..113S}
}

@article{2024ApJ...962..141C,
  author        = {{Chan}, Man Ho and {Lee}, Chak Man and {Cui}, Lang and {Chang}, Ning and {Leung}, Chun Sing},
  title         = {{Constraining Annihilating Dark Matter Using the Multifrequency Radio Flux Profiles of the M33 Galaxy}},
  journal       = {\apj},
  year          = 2024,
  month         = feb,
  volume        = {962},
  number        = {2},
  eid           = {141},
  pages         = {141},
  doi           = {10.3847/1538-4357/ad2172},
  archiveprefix = {arXiv},
  eprint        = {2401.11668},
  primaryclass  = {astro-ph.GA},
  adsurl        = {https://ui.adsabs.harvard.edu/abs/2024ApJ...962..141C}
}

@article{2024ApJ...968...46D,
  author        = {{De la Torre Luque}, Pedro and {Balaji}, Shyam and {Koechler}, Jordan},
  title         = {{Importance of Cosmic-Ray Propagation on Sub-GeV Dark Matter Constraints}},
  journal       = {\apj},
  year          = 2024,
  month         = jun,
  volume        = {968},
  number        = {1},
  eid           = {46},
  pages         = {46},
  doi           = {10.3847/1538-4357/ad41e0},
  archiveprefix = {arXiv},
  eprint        = {2311.04979},
  primaryclass  = {hep-ph},
  adsurl        = {https://ui.adsabs.harvard.edu/abs/2024ApJ...968...46D}
}

@article{2024ApJ...973L..14D,
  author        = {{DES Collaboration} and {Abbott}, T.~M.~C. and {Acevedo}, M. and {Aguena}, M. and {Alarcon}, A. and {Allam}, S. and {Alves}, O. and {Amon}, A. and {Andrade-Oliveira}, F. and {Annis}, J. and {Armstrong}, P. and {Asorey}, J. and {Avila}, S. and {Bacon}, D. and {Bassett}, B.~A. and {Bechtol}, K. and {Bernardinelli}, P.~H. and {Bernstein}, G.~M. and {Bertin}, E. and {Blazek}, J. and {Bocquet}, S. and {Brooks}, D. and {Brout}, D. and {Buckley-Geer}, E. and {Burke}, D.~L. and {Camacho}, H. and {Camilleri}, R. and {Campos}, A. and {Carnero Rosell}, A. and {Carollo}, D. and {Carr}, A. and {Carretero}, J. and {Castander}, F.~J. and {Cawthon}, R. and {Chang}, C. and {Chen}, R. and {Choi}, A. and {Conselice}, C. and {Costanzi}, M. and {da Costa}, L.~N. and {Crocce}, M. and {Davis}, T.~M. and {DePoy}, D.~L. and {Desai}, S. and {Diehl}, H.~T. and {Dixon}, M. and {Dodelson}, S. and {Doel}, P. and {Doux}, C. and {Drlica-Wagner}, A. and {Elvin-Poole}, J. and {Everett}, S. and {Ferrero}, I. and {Fert{\'e}}, A. and {Flaugher}, B. and {Foley}, R.~J. and {Fosalba}, P. and {Friedel}, D. and {Frieman}, J. and {Frohmaier}, C. and {Galbany}, L. and {Garc{\'\i}a-Bellido}, J. and {Gatti}, M. and {Gaztanaga}, E. and {Giannini}, G. and {Glazebrook}, K. and {Graur}, O. and {Gruen}, D. and {Gruendl}, R.~A. and {Gutierrez}, G. and {Hartley}, W.~G. and {Herner}, K. and {Hinton}, S.~R. and {Hollowood}, D.~L. and {Honscheid}, K. and {Huterer}, D. and {Jain}, B. and {James}, D.~J. and {Jeffrey}, N. and {Kasai}, E. and {Kelsey}, L. and {Kent}, S. and {Kessler}, R. and {Kim}, A.~G. and {Kirshner}, R.~P. and {Kovacs}, E. and {Kuehn}, K. and {Lahav}, O. and {Lee}, J. and {Lee}, S. and {Lewis}, G.~F. and {Li}, T.~S. and {Lidman}, C. and {Lin}, H. and {Malik}, U. and {Marshall}, J.~L. and {Martini}, P. and {Mena-Fern{\'a}ndez}, J. and {Menanteau}, F. and {Miquel}, R. and {Mohr}, J.~J. and {Mould}, J. and {Muir}, J. and {M{\"o}ller}, A. and {Neilsen}, E. and {Nichol}, R.~C. and {Nugent}, P. and {Ogando}, R.~L.~C. and {Palmese}, A. and {Pan}, Y.-C. and {Paterno}, M. and {Percival}, W.~J. and {Pereira}, M.~E.~S. and {Pieres}, A. and {Malag{\'o}n}, A.~A. Plazas and {Popovic}, B. and {Porredon}, A. and {Prat}, J. and {Qu}, H. and {Raveri}, M. and {Rodr{\'\i}guez-Monroy}, M. and {Romer}, A.~K. and {Roodman}, A. and {Rose}, B. and {Sako}, M. and {Sanchez}, E. and {Sanchez Cid}, D. and {Schubnell}, M. and {Scolnic}, D. and {Sevilla-Noarbe}, I. and {Shah}, P. and {Smith}, J. Allyn. and {Smith}, M. and {Soares-Santos}, M. and {Suchyta}, E. and {Sullivan}, M. and {Suntzeff}, N. and {Swanson}, M.~E.~C. and {S{\'a}nchez}, B.~O. and {Tarle}, G. and {Taylor}, G. and {Thomas}, D. and {To}, C. and {Toy}, M. and {Troxel}, M.~A. and {Tucker}, B.~E. and {Tucker}, D.~L. and {Uddin}, S.~A. and {Vincenzi}, M. and {Walker}, A.~R. and {Weaverdyck}, N. and {Wechsler}, R.~H. and {Weller}, J. and {Wester}, W. and {Wiseman}, P. and {Yamamoto}, M. and {Yuan}, F. and {Zhang}, B. and {Zhang}, Y.},
  title         = {{The Dark Energy Survey: Cosmology Results with {\ensuremath{\sim}}1500 New High-redshift Type Ia Supernovae Using the Full 5 yr Data Set}},
  journal       = {\apjl},
  year          = 2024,
  month         = sep,
  volume        = {973},
  number        = {1},
  eid           = {L14},
  pages         = {L14},
  doi           = {10.3847/2041-8213/ad6f9f},
  archiveprefix = {arXiv},
  eprint        = {2401.02929},
  primaryclass  = {astro-ph.CO},
  adsurl        = {https://ui.adsabs.harvard.edu/abs/2024ApJ...973L..14D}
}

@article{2024arXiv241201994M,
  author        = {{Moresco}, Michele},
  title         = {{Measuring the expansion history of the Universe with cosmic chronometers}},
  journal       = {arXiv e-prints},
  year          = 2024,
  month         = dec,
  eid           = {arXiv:2412.01994},
  pages         = {arXiv:2412.01994},
  doi           = {10.48550/arXiv.2412.01994},
  archiveprefix = {arXiv},
  eprint        = {2412.01994},
  primaryclass  = {astro-ph.CO},
  adsurl        = {https://ui.adsabs.harvard.edu/abs/2024arXiv241201994M}
}

@article{2024EPJST.233.2113D,
  author  = {{Dutta}, Manoranjan and {Mahapatra}, Satyabrata},
  title   = {{Mini-review on self-interacting dark matter}},
  journal = {European Physical Journal Special Topics},
  year    = 2024,
  month   = oct,
  volume  = {233},
  number  = {11-12},
  pages   = {2113-2128},
  doi     = {10.1140/epjs/s11734-024-01121-6},
  adsurl  = {https://ui.adsabs.harvard.edu/abs/2024EPJST.233.2113D}
}

@article{2024MNRAS.528.3433S,
  author        = {{Siegert}, Thomas and {Calore}, Francesca and {Serpico}, Pasquale Dario},
  title         = {{Sub-GeV dark matter annihilation: limits from Milky Way observations with INTEGRAL}},
  journal       = {\mnras},
  year          = 2024,
  month         = feb,
  volume        = {528},
  number        = {2},
  pages         = {3433-3440},
  doi           = {10.1093/mnras/stae104},
  archiveprefix = {arXiv},
  eprint        = {2401.03795},
  primaryclass  = {astro-ph.HE},
  adsurl        = {https://ui.adsabs.harvard.edu/abs/2024MNRAS.528.3433S}
}

@article{2024PhRvD.110l3026Z,
  author        = {{Zakharov}, E.~I. and {Barinov}, V.~V. and {Burenin}, R.~A. and {Gorbunov}, D.~S. and {Krivonos}, R.~A.},
  title         = {{Constraints on the parameters of keV-scale mass annihilating dark matter obtained with SRG/ART-XC observations}},
  journal       = {\prd},
  year          = 2024,
  month         = dec,
  volume        = {110},
  number        = {12},
  eid           = {123026},
  pages         = {123026},
  doi           = {10.1103/PhysRevD.110.123026},
  archiveprefix = {arXiv},
  eprint        = {2407.18371},
  primaryclass  = {astro-ph.HE},
  adsurl        = {https://ui.adsabs.harvard.edu/abs/2024PhRvD.110l3026Z}
}

@article{2025ApJ...978L..33G,
  author        = {{Guo}, Wuzheng and {Wang}, Qiumin and {Cao}, Shuo and {Biesiada}, Marek and {Liu}, Tonghua and {Lian}, Yujie and {Jiang}, Xinyue and {Mu}, Chengsheng and {Cheng}, Dadian},
  title         = {{Newest Measurements of Hubble Constant from DESI 2024 Baryon Acoustic Oscillation Observations}},
  journal       = {\apjl},
  year          = 2025,
  month         = jan,
  volume        = {978},
  number        = {2},
  eid           = {L33},
  pages         = {L33},
  doi           = {10.3847/2041-8213/ada37f},
  archiveprefix = {arXiv},
  eprint        = {2412.13045},
  primaryclass  = {astro-ph.CO},
  adsurl        = {https://ui.adsabs.harvard.edu/abs/2025ApJ...978L..33G}
}

@article{2025ApJ...981L..24L,
  author        = {{Liu}, Tonghua and {Wang}, Shengjia and {Wu}, Hengyu and {Cao}, Shuo and {Wang}, Jieci},
  title         = {{Newest Measurements of Cosmic Curvature with BOSS/eBOSS and DESI DR1 Baryon Acoustic Oscillation Observations}},
  journal       = {\apjl},
  year          = 2025,
  month         = mar,
  volume        = {981},
  number        = {2},
  eid           = {L24},
  pages         = {L24},
  doi           = {10.3847/2041-8213/adb7de},
  archiveprefix = {arXiv},
  eprint        = {2411.14154},
  primaryclass  = {astro-ph.CO},
  adsurl        = {https://ui.adsabs.harvard.edu/abs/2025ApJ...981L..24L}
}

\end{document}